\documentclass[conference,compsoc]{IEEEtran}
\ifCLASSOPTIONcompsoc
  \usepackage[nocompress]{cite}
\else
  \usepackage{cite}
\fi
\ifCLASSINFOpdf
\else
\fi

\usepackage{amsmath,amsfonts,amsthm}
\usepackage{algorithmic}
\usepackage{algorithm}
\usepackage{booktabs}
\usepackage{array}
\usepackage{multirow}
\usepackage{textcomp}
\usepackage{stfloats}
\usepackage{url}
\usepackage{verbatim}
\usepackage{graphicx}
\usepackage{cite}
\usepackage{amssymb}
\usepackage{tabularray} % Load the tabularray package
\usepackage{textcomp}
\usepackage{stfloats}
\usepackage{url}
\usepackage{color}
\usepackage{verbatim}
\usepackage{graphicx}
\usepackage{cite}

\theoremstyle{definition}
\newtheorem{definition}{\textbf{Definition}}

\usepackage{hyperref}
\usepackage[normalem]{ulem}
\usepackage{multirow}
\usepackage{enumitem}
\usepackage{xspace}
\newcommand{\etal}{et al.\xspace}

\newcommand{\ie}{i.e.,\xspace}

\usepackage{marvosym}

\newcommand{\graph}{HAMG\xspace}
\newcommand{\model}{\textit{LMDetect}\xspace}

\begin{document}
\title{Lateral Movement Detection via Time-aware Subgraph Classification on Authentication Logs}

\author{\IEEEauthorblockN{Anonymous authors}}

% \author{\IEEEauthorblockN{Michael Shell}
% \IEEEauthorblockA{School of Electrical and\\Computer Engineering\\
% Georgia Institute of Technology\\
% Atlanta, Georgia 30332--0250\\
% Email: http://www.michaelshell.org/contact.html}
% \and
% \IEEEauthorblockN{Homer Simpson}
% \IEEEauthorblockA{Twentieth Century Fox\\
% Email: homer@thesimpsons.com}
% \and
% \IEEEauthorblockN{James Kirk\\ and Montgomery Scott}
% \IEEEauthorblockA{Starfleet Academy\\
% San Francisco, California 96678-2391\\
% Telephone: (800) 555--1212\\


% compsoc conferences compared to traditional conferences), use this

% 共： \textsuperscript{\textsection}

\author{
    \IEEEauthorblockN{Jiajun Zhou\IEEEauthorrefmark{1}\IEEEauthorrefmark{2}\IEEEauthorrefmark{3},
                      Jiacheng Yao\IEEEauthorrefmark{1}\IEEEauthorrefmark{2},
                      Xuanze Chen\IEEEauthorrefmark{1}\IEEEauthorrefmark{2}\IEEEauthorrefmark{3}, 
                      Shanqing Yu\IEEEauthorrefmark{1}\IEEEauthorrefmark{2},
                      Qi Xuan\IEEEauthorrefmark{1}\IEEEauthorrefmark{2}\textsuperscript{\Letter}, and
                      Xiaoniu Yang\IEEEauthorrefmark{1}\IEEEauthorrefmark{4}
                      }
\IEEEauthorblockA{\IEEEauthorrefmark{1}Institute of Cyberspace Security, Zhejiang University of Technology, Hangzhou 310023, China}
\IEEEauthorblockA{\IEEEauthorrefmark{2}Binjiang Institute of Artificial Intelligence, ZJUT, Hangzhou 310056, China}
\IEEEauthorblockA{\IEEEauthorrefmark{3}College of Computer Science and Technology, Zhejiang University of Technology,  Hangzhou 310023, China}
\IEEEauthorblockA{\IEEEauthorrefmark{4}Science and Technology on Communication Information Security Control Laboratory, Jiaxing 314033, China}
Email: \{jjzhou, yaojc, chenxuanze, yushanqing, xuanqi\}@zjut.edu.cn
% \thanks{\IEEEauthorrefmark{*}Qi Xuan is the corresponding author (email: xuanqi@zjut.edu.cn)}.
}


\maketitle

\begingroup
\renewcommand\thefootnote{\Letter}\footnotetext{Corresponding author}
\endgroup

% \begingroup
% \renewcommand\thefootnote{\Letter}\footnotetext{Corresponding author}
% \endgroup

\begin{abstract}
% （修改后）横向移动是网络高级持续性威胁攻击中关键的一环，攻击者通过利用内网和物联网设备的安全漏洞，在成功渗透后扩展控制范围，以窃取敏感数据或实施其他恶意活动，对系统安全构成了严重威胁。现有的研究指出，攻击者通常会利用看似无关的操作来掩盖其恶性意图，以此规避现有的横向移动检测手段，隐藏其入侵轨迹。对此，我们从图的视角分析主机认证日志数据，并提出了一个多尺度横向移动检测框架，叫做LMDetect。该框架的主要工作流程如下:1) 将主机认证日志数据构建为一个异构图来加强系统内部实体的关联性；2) 利用时间感知的子图生成器从异构认证图中获取以身份认证事件为中心的子图；2）设计了一个多尺度注意力编码器，利用局部和全局注意力来捕获认证子图中隐藏的异常行为模式，实现横向移动检测。在两个真实的认证日志数据集上的大量实验表明了我们的框架在检测横向移动行为的有效性和优越性。
Lateral movement is a crucial component of advanced persistent threat (APT) attacks in networks. Attackers exploit security vulnerabilities in internal networks or IoT devices, expanding their control after initial infiltration to steal sensitive data or carry out other malicious activities, posing a serious threat to system security. Existing research suggests that attackers generally employ seemingly unrelated operations to mask their malicious intentions, thereby evading existing lateral movement detection methods and hiding their intrusion traces. In this regard, we analyze host authentication log data from a graph perspective and propose a multi-scale lateral movement detection framework called \textit{LMDetect}. The main workflow of this framework proceeds as follows: 1) Construct a heterogeneous multigraph from host authentication log data to strengthen the correlations among internal system entities; 2) Design a time-aware subgraph generator to extract subgraphs centered on authentication events from the heterogeneous authentication multigraph; 3) Design a multi-scale attention encoder that leverages both local and global attention to capture hidden anomalous behavior patterns in the authentication subgraphs, thereby achieving lateral movement detection. Extensive experiments on two real-world authentication log datasets demonstrate the effectiveness and superiority of our framework in detecting lateral movement behaviors.
\end{abstract}


% \fi
\IEEEpeerreviewmaketitle

\section{Introduction}
% 近年来，互联网的迅猛发展深刻地改变了数字化环境，带来了前所未有的连接性和信息获取途径。然而，这种高度连通性也引发了一系列重大的网络安全挑战。特别是高级持续性威胁（Advanced Persistent Threat，APT）的复杂性和隐蔽性不断提高，对互联网安全构成了严峻的威胁。在APT攻击中，横向移动攻击尤为突出，已成为网络安全领域亟待解决的重要问题。横向移动是APT攻击生命周期中的关键阶段。攻击者在获得网络中某个节点的初始访问权限后，利用多种技术手段，在网络内部逐步渗透和扩展控制范围。这一过程使得攻击者能够提升权限、访问敏感数据，并建立持久化的访问机制（如后门、恶意软件等）。典型的横向移动场景包括：攻击者通过钓鱼邮件或漏洞利用入侵一台主机后，扫描网络以识别其他设备的漏洞并逐步渗透；或内部用户利用其合法权限或窃取的凭据执行未经授权的操作，从而扩大对系统的控制范围。这些行为的最终目的是窃取高价值的资源，会给组织带来严重的损失。因此，检测横向移动行为对于阻止APT攻击至关重要。

\IEEEPARstart{R}{e}cently, the rapid development of the internet has profoundly transformed the digital environment, introducing unprecedented connectivity and avenues for information access. However, this high degree of connectivity has also triggered a series of significant cybersecurity challenges. In particular, the complexity and stealthiness of Advanced Persistent Threats (APTs) have escalated, posing severe threats to internet security. Among APT attacks, lateral movement attacks are especially prominent and have become a critical issue that urgently needs to be addressed in cybersecurity.

\begin{figure*}[!htb] 
    \centering 
    \includegraphics[width=1\textwidth]{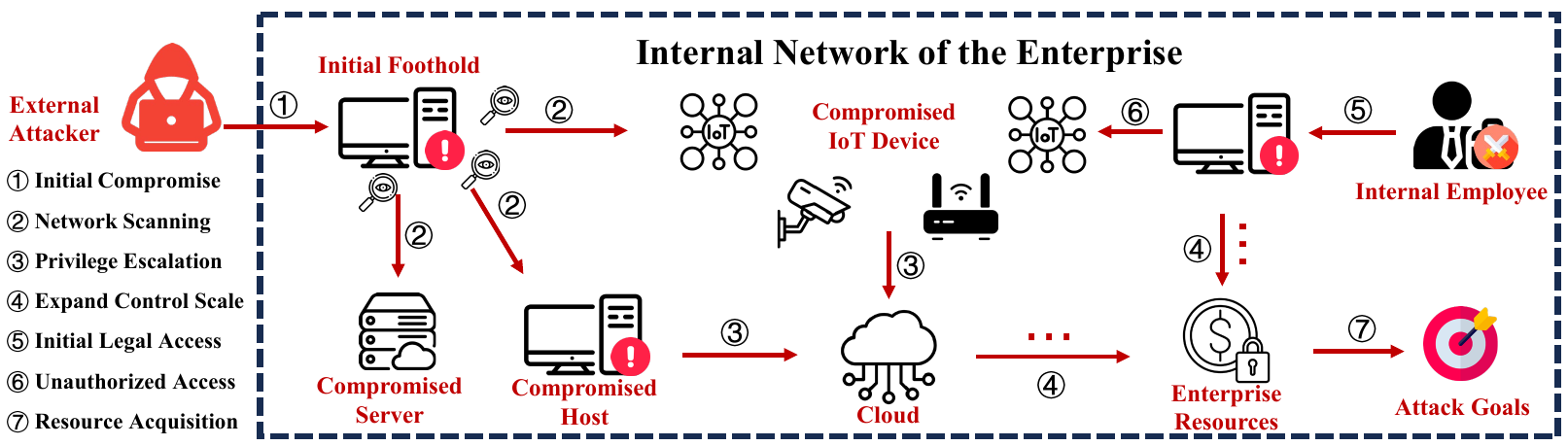} 
    \caption{Two lateral movement scenarios in the enterprise internal network: 1) External threat actors employ advanced persistent threat (APT) techniques to infiltrate the internal network; 2) Internal personnel exploit initial privileges for unauthorized access. Both leverage lateral movement tactics to expand their control and accomplish the objective of exfiltrating sensitive data.} 
\label{fig: LM process}
\end{figure*}

Lateral movement is a key stage in the APT attack. After gaining initial access to a node within the network, attackers employ various techniques to progressively infiltrate and expand their control within the network. This process allows attackers to escalate privileges, access sensitive data, and establish persistent access mechanisms (such as backdoors or malware). 
Fig~\ref{fig: LM process} illustrates two typical lateral movement scenarios in an internal network: 
1) External attackers successfully infiltrate an internal network host through phishing emails or vulnerability exploitation. After establishing an initial foothold, they gradually expand their control by scanning other devices in the network and exploiting inherent security weaknesses or by stealing credentials to escalate privileges;
2) Internal members leverage their initial legitimate access or stolen credentials to carry out unauthorized access, progressively extending their control over the system.
% Typical lateral movement scenarios include attackers infiltrating a host via phishing emails or exploiting vulnerabilities, then scanning the network to identify vulnerabilities in other devices and gradually penetrating them; or internal users leveraging their legitimate privileges or stolen credentials to execute unauthorized operations, thereby expanding their control over the system. 
The ultimate goal of lateral movement is to steal high-value resources, which can lead to significant losses for organizations. Therefore, detecting lateral movement behavior is critical to thwarting APT attacks.

% 对此，研究者们开发了多种横向移动检测方法，主要包括基于端点检测与响应（Endpoint Detection and Response，EDR）的方法、机器学习方法、深度学习方法。这些方法虽然在检测横向移动方面取得了一定成效，但是依然存在各自的局限性。基于端点检测与响应（Endpoint Detection and Response，EDR）的方法虽然能够实时监控端点设备并检测异常行为，但容易被攻击者利用合法工具和凭证进行规避，从而绕过检测。此外，EDR需要大量计算资源来处理海量数据，可能导致性能瓶颈。机器学习方法在识别复杂行为模式方面具有较好的表现，但它们依赖大量的标记数据进行训练，而攻击者不断变换策略，导致现有模型难以适应新的攻击手法。另外，机器学习模型可能产生误报，增加运维负担。深度学习方法在提取数据中的复杂特征方面具有优势，但同样需要大量高质量的数据进行训练，并且训练过程复杂、耗时。此外，深度学习模型的可解释性较差，难以分析和理解检测结果，给安全专家的威胁响应带来挑战。
Researchers have developed various methods for detecting lateral movement, primarily including methods based on endpoint detection and response (EDR)~\cite{noureddine2016game,khoury2020hybrid,nassar2021game,sawilla2008identifying,bowen2009baiting,ho2021hopper,han2020unicorn,milajerdi2019holmes}, machine learning (ML) and deep learning (DL)~\cite{bai2021rdp,meijerink2019anomaly,smiliotopoulos2023detection,bian2021uncovering,he2023comprehensive,bai2019machine,sarker2021cyberlearning,parra2019implementation,parra2020detecting}. While these approaches have achieved certain success in detecting lateral movement, they still face several limitations. EDR-based methods can monitor endpoint devices in real-time and detect abnormal behaviors, but they are vulnerable to attackers who can evade detection by using legitimate tools and credentials. Additionally, EDR requires substantial computational resources to process large volumes of data, potentially leading to performance bottlenecks. ML-based methods demonstrate good performance in identifying complex behavioral patterns, but they rely heavily on large amounts of labeled data for training. As attackers continuously modify their  tactics, existing models struggle to adapt to new attack means. DL-based methods excel in extracting complex features from data but also require extensive high-quality data for training, and the training process is complex and time-consuming. Moreover, DL-based models have poor interpretability, making it difficult to analyze and understand the detection results, thus posing challenges for security experts in threat response.

% 作为检测横向移动的重要数据资源，身份认证日志记录了用户、设备和服务之间的每一次身份验证尝试，包括成功和失败的登录，以及相关的时间、位置和上下文信息。这些数据为安全分析师和自动化系统提供了对内部网络活动的微观视角，能够追踪和分析用户与设备的行为模式。近期，基于身份认证日志的图神经网络（Graph Neural Network，GNN）方法在横向移动检测领域取得了显著成效。通过将身份认证日志建模为异构图结构，并利用GNN学习横向移动行为的高阶表示，这些方法能够捕捉横向移动的复杂时序和关系特征，取得了优异的检测性能。然而，基于GNN的方法依然存在特定的局限性：1）检测范式的局限性。现有的GNN方法通常将横向移动检测建模为基于节点表征学习的边分类任务，孤立地看待网络行为，忽略了横向移动通常涉及一系列连续行为的事实；2）大规模图计算的效率限制。身份验证日志数据规模庞大，记录了大量的用户认证事件和网络交互。在如此大规模的网络交互图上进行全图学习会导致计算复杂度过高和时间成本过大；3）身份验证日志数据的不平衡分布。在典型的网络环境中，良性交互数据占据绝大多数，而横向移动为代表的恶意数据相对稀少，导致正负样本比例极度不平衡。现有基于GNN的检测模型在此情况下容易产生检测偏差，倾向于预测常见的良性行为，忽略稀有但危害巨大的恶意行为。这种数据不平衡性对模型的准确性和鲁棒性提出了严峻的挑战。
Authentication logs serve as a critical data resource for detecting lateral movements, recording every authentication attempt between users, devices, and services, including both successful and failed logins, along with related timestamps, locations, and contextual information. These data provide security analysts and automated systems with a microscopic view of internal network activities, enabling the tracking and analysis of user and device behavioral patterns. Recently, methods based on Graph Neural Networks (GNNs)~\cite{liu2018latte,king2023euler,fang2022lmtracker,bowman2020detecting,liu2019log2vec} utilizing authentication logs have achieved notable success in the domain of lateral movement detection. By modeling authentication logs as heterogeneous graph structures and employing GNNs to learn higher-order representations of lateral movement behaviors, these methods have captured complex temporal and relational characteristics of lateral movements, resulting in superior detection performance. However, GNN-based approaches still face specific limitations: 
1) \textbf{Limitations of the detection paradigm.} Current GNN-based methods typically model lateral movement detection as an edge classification task based on node representation learning, treating network behaviors in isolation and overlooking the fact that lateral movements often involve a series of continuous actions; 
2) \textbf{Efficiency constraints of large-scale graph computation.} The scale of authentication log data is vast, recording numerous user authentication events and network interactions. Learning on such large-scale interaction graphs results in high computational complexity and significant time costs; 
3) \textbf{Imbalanced distribution of authentication log data.} In typical network environments, benign interactions vastly outnumber malicious ones, such as lateral movements, leading to a severe imbalance between positive and negative samples. Existing GNN-based detection models are prone to detection bias under these conditions, favoring predictions of common benign behaviors and overlooking rare but highly damaging malicious actions. This imbalance poses serious challenges to the accuracy and robustness of models.

\begin{figure}[!htb] 
    \centering 
    \includegraphics[width=\linewidth]{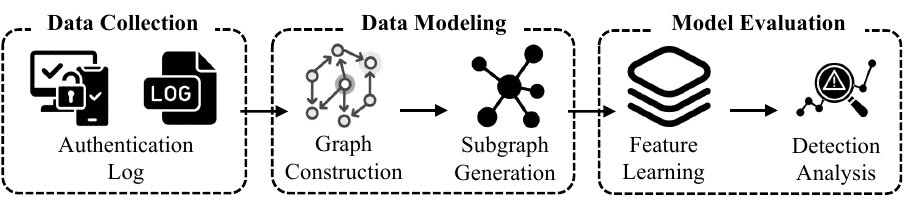} 
    \caption{Workflow of the \model framework.} 
\label{fig: workflow}
\end{figure}
% 对此，我们提出了一个时间感知的多尺度横向移动检测框架，一个端到端的图神经网络模型，来刻画网络认证事件的行为模式并深远地实现横向移动检测。具体而言，我们首先解析认证日志数据并归纳网络实体和交互类型，构建可以描述网络活动的异构认证多图。然后，我们将横向移动检测建模为子图层面的分类任务，并设计了一个时间感知子图生成器来捕捉目标认证事件周围的高相关信息，形成以目标认证事件为中心的时间感知认证子图，它允许模型以小批量的方式训练来提高效率。为了更好地刻画认证事件的行为模式，我们进一步设计了一个多尺度注意力编码器作为我们框架的骨干，它可以有效地学习网络实体间的局部和全局依赖，实现对认证事件的强大表征。本工作的主要贡献可以总结如下：
% 1) 数据和任务建模：我们将认证日志建模为异构认证多图，并将横向移动检测问题建模为子图分类任务，设计了时间感知子图生成器来实现可扩展的横向移动检测。
% 2）强大的表征能力：我们提出了一个多尺度注意力编码器来有效地学习网络认证活动中的局部和全局依赖，捕捉认证事件的行为模式。
% 3）sota的性能：在两个真实的认证日志数据集上的大量实验表明了我们框架在横向移动检测任务上的有效性和显著的优越性。
In this regard, we propose a time-aware multi-scale \textbf{L}ateral \textbf{M}ovement \textbf{Detect}ion framework (\model) — an end-to-end graph neural network model — to characterize behavior patterns in network authentication events and further achieve lateral movement detection. Figure~\ref{fig: workflow} illustrates the workflow of \model.
Specifically, we first parse authentication log data, identify network entities and interaction types, and construct a heterogeneous authentication multigraph (\graph) to describe network activity. 
We then consider lateral movement detection as a subgraph-level classification problem, designing a time-aware subgraph generator to capture highly relevant information surrounding target authentication events, yielding time-aware authentication subgraphs centered on each target event. This approach enables efficient mini-batch training of \model. 
To better capture the behavior patterns of authentication events, we further introduce a multi-scale attention encoder as the backbone of our framework, which can effectively learn both local and global dependencies among network entities, providing strong representations of authentication events. The main contributions of this work are as follows:
\begin{itemize}[leftmargin=10pt]
    \item \textbf{Data and Task Modeling:} We model authentication logs as a heterogeneous authentication multigraph and formulate the lateral movement detection task as a subgraph classification problem, designing a time-aware subgraph generator to enable scalable lateral movement detection.
    \item \textbf{Powerful Representation Capability:} We propose a multi-scale attention encoder that effectively learns both local and global dependencies in network activities, capturing the behavior patterns of authentication events.
    \item \textbf{State-of-the-Art Performance:} Extensive experiments on two real-world authentication log datasets demonstrate the effectiveness and significant superiority of our framework for the lateral movement detection task.
\end{itemize}

\section{Related work}
\subsection{Traditional Lateral Movement Detection}
% 早期的横向移动检测方法主要依赖网络配置和漏洞数据，通过识别关键资产并分析日志来检测潜在的横向移动路径。然而，这些方法在应对复杂攻击策略时存在局限性，因此在实际应用中面临挑战。为了克服这些局限性，研究者们逐步对这些方法进行了扩展和改进。
Early lateral movement detection methods primarily relies on network configuration and vulnerability data, detecting potential lateral movement paths by identifying key assets and analyzing logs. However, these methods face limitations in addressing complex attack strategies, posing challenges for practical application. To overcome these limitations, researchers have gradually expanded and improved these approaches.

% 例如，Noureddine 和 Nassar 等人 \cite{noureddine2016game, nassar2021game} 提出了基于博弈论的方法，通过动态建模攻击者和防御者之间的互动，分析攻击者的横向移动路径，并评估防御策略的有效性。然而，这些方法假设攻击者和防御者的行为模式是固定的，未考虑攻击者的多步策略及其对防御措施的适应性，从而限制了模型的实用性和泛化能力。另外，有研究通过攻击图实现检测，Sawilla 等人 \cite{sawilla2008identifying} 提出了 AssetRank 算法，通过对依赖攻击图的节点进行排序，识别关键攻击资产。然而，该方法依赖准确的网络配置和漏洞数据，并需要专业知识来设置参数和选择向量。Hopper 系统 \cite{ho2021hopper} 基于企业日志构建登录活动图，通过检测规则和异常评分算法来识别横向移动路径，取得了较高的检测率和较低的误报率，但其性能受网络架构、权限管理策略和日志准确性的影响。Milajerdi 等人 \cite{milajerdi2019holmes} 提出的 HOLMES 系统通过构建高层次情境图，并利用主机审计数据关联APT中的信息流来检测横向移动。然而，该方法可能错过那些没有直接系统调用信息流的横向移动。Bai 等人 \cite{bai2021rdp} 基于Windows远程桌面协议（RDP）事件日志提出了基于机器学习的方法，通过监督学习技术和检测规则结合新型异常评分算法，能够有效识别与横向移动相关的RDP会话，但该方法依赖特定数据集提取的特征，需要根据不同环境或攻击模式进行适配和重新训练。Smiliotopoulos 等人 \cite{smiliotopoulos2023detection} 基于Sysmon日志提出了一种综合的横向移动检测方法，结合监督机器学习和特征选择技术，在Windows平台上取得了高F1分数和AUC分数，但该方法可能不适用于除Windows之外的其他操作系统。

For instance, Noureddine and Nassar \etal~\cite{noureddine2016game,nassar2021game} proposed game-theoretic methods to dynamically model interactions between attackers and defenders, analyzing attackers' lateral movement paths and evaluating the effectiveness of defense strategies. However, these methods assume fixed behavior patterns for both attackers and defenders, without considering attackers' multi-step strategies and adaptability to defenses, which limits the models' practicality and generalization capabilities.
Other studies leverage attack graphs for detection. Sawilla \etal~\cite{sawilla2008identifying} proposed the AssetRank algorithm, which ranks nodes in a dependency attack graph to identify critical attack assets. However, this method depends on precise network configuration and vulnerability data and requires expert knowledge for parameter settings and vector selection. The Hopper system~\cite{ho2021hopper} constructs a login activity graph from enterprise logs, identifying lateral movement paths through detection rules and anomaly scoring algorithms, achieving high detection rates and low false positive rates. However, its performance is influenced by network architecture, permission management policies, and log accuracy.
Milajerdi \etal~\cite{milajerdi2019holmes} developed the HOLMES system, which constructs high-level contextual graphs and uses host audit data to correlate information flows in APTs for lateral movement detection. However, this approach may miss lateral movement activities that lack direct system call information flow. Bai \etal~\cite{bai2021rdp} proposed a machine learning-based method utilizing Windows Remote Desktop Protocol (RDP) event logs, combining supervised learning with detection rules and a novel anomaly scoring algorithm to effectively identify RDP sessions related to lateral movement. However, this method depends on features extracted from a specific dataset, requiring adaptation and retraining for different environments or attack patterns.
Smiliotopoulos \etal~\cite{smiliotopoulos2023detection} introduced a comprehensive lateral movement detection approach based on Sysmon logs, combining supervised machine learning with feature selection techniques, achieving high F1 and AUC scores on the Windows platform. However, this method may not generalize well to operating systems other than Windows.

\subsection{Lateral Movement Detection Based on GNNs}
% 近年来，图神经网络（GNNs）在多个领域，如欺诈检测 \cite{jin2022heterogeneous, zhou2022behavior}、网络安全 和 推荐系统 \cite{he2020lightgcn}，展现了强大的表示学习能力，其在横向移动检测领域的应用也日益增多。研究者通过构建网络服务图、认证图和异构认证日志图，利用GNNs捕捉和分析网络中复杂的关系和动态变化，以实现检测。
In recent years, Graph Neural Networks (GNNs) have demonstrated strong representation learning capabilities across various fields, such as fraud detection~\cite{jin2022heterogeneous,zhou2022behavior,10490264}, cybersecurity~\cite{bilot2023graph}, and social analysis~\cite{zhou2024pathmlp,zhou2023clarify}. Their applications in the domain of lateral movement detection are also increasing. Researchers construct network service graphs, authentication graphs, and heterogeneous authentication log graphs, leveraging GNNs to capture and analyze complex relationships and dynamic changes within the network to achieve detection.

%例如，Liu 等人 \cite{liu2018latte} 提出的 Latte 系统通过已知感染节点的组合分析和取证，快速识别感染计算机，并利用远程文件执行检测器过滤稀有路径，识别未知的横向移动攻击。尽管该系统在大规模数据集上展现了高效性和有效性，但仍面临区分攻击者和系统管理员正常操作的挑战。Parra 等人 \cite{parra2020detecting} 提出了一个分布式深度学习框架，用于检测钓鱼攻击和应用层DDoS攻击，框架结合了IoT设备中的DCNN和基于时间的LSTM模型，能够有效识别跨设备分布式钓鱼攻击。Fang 等人 \cite{fang2022lmtracker} 提出的 LMTracker 构建了一个基于认证日志的异构图，通过抽象横向移动为路径，使用预定义阈值和元路径进行无监督异常路径检测。然而，该方法未考虑时间信息，而时间信息对于理解攻击者的横向移动序列至关重要。Liu 等人 \cite{liu2019log2vec} 通过特定规则建模用户行为，并应用聚类方法将相似日志条目分组，通过聚类大小和数量来判断恶意行为。King 等人 \cite{king2023euler} 提出的 Euler 系统则通过生成每个时间段的离散动态图快照，并利用GNNs编码不断变化的拓扑结构，将输出通过RNN等顺序编码器处理，以捕捉网络动态。但该方法存在较高的计算开销。
For example, Liu \etal~\cite{liu2018latte} proposed the Latte system, which quickly identifies infected computers through combined analysis and forensics of known compromised nodes, using a remote file execution detector to filter rare paths and identify unknown lateral movement attacks. 
Although this system has demonstrated efficiency and effectiveness on large-scale datasets, it still faces challenges in distinguishing between attacker activity and normal actions by system administrators. 
Parra \etal~\cite{parra2020detecting} proposed a distributed deep learning framework for detecting phishing and application-layer DDoS attacks. The framework combines DCNNs in IoT devices with time-based LSTM models, effectively identifying cross-device distributed phishing attacks. 
Fang \etal~\cite{fang2022lmtracker} developed LMTracker, which constructs a heterogeneous graph based on authentication logs, abstracts lateral movement as a path, and uses predefined thresholds and meta-paths for unsupervised anomalous path detection. However, this approach does not consider temporal information, which is crucial for understanding the sequence of lateral movement by attackers. 
Liu \etal~\cite{liu2019log2vec} used specific rules to model user behavior and applied clustering methods to group similar log entries, determining malicious behavior based on cluster size and quantity. 
King \etal~\cite{king2023euler} proposed the Euler system, which generates discrete dynamic graph snapshots for each time period, encoding the evolving topology using GNNs, and processes the outputs with sequential encoders such as recurrent neural network (RNNs) to capture network dynamics. However, this approach incurs significant computational overhead.

%虽然基于GNNs的横向移动检测方法已取得了一定进展，但大多数研究集中在实体之间的单一连接上，这可能导致忽视更复杂的网络行为模式。横向移动是一个持续的过程，涉及一系列操作，单纯通过边分类或链路预测来评估网络活动的状态可能无法揭示完整的攻击路径。
% While GNN-based methods for lateral movement detection have achieved some progress, most studies focus on single connections between entities, which may overlook more complex patterns of network behavior. Lateral movement is an ongoing process involving a series of operations, and simply assessing network activity through edge classification or link prediction may not fully reveal the complete attack path.

Although GNN-based approaches have made some progress in detecting lateral movement, most studies primarily focus on individual connections between entities, potentially overlooking the existence of more intricate patterns in network behavior. Lateral movement is an ongoing process that involves a series of operations, and solely evaluating network activity via edge classification or link prediction may not fully uncover the complete attack path.

\section{Heterogeneous Authentication Multigraph}
% In this work, we will construct a Heterogeneous Authentication Graph (HAG) for authentication logs and detect lateral movement behaviors from a graph classification perspective. In this section, we first model the proposed target problem, then describe the construction of the HAG in detail, and finally the temporal interaction mapping $TempIntMap$ is defined to lightly incorporate temporal information into our model.

\begin{table}[!htb]
    \renewcommand\arraystretch{1.2} 
    \centering
    \caption{Information of Authentication Logs.}
    \resizebox{\linewidth}{!}{
    \begin{tabular}{lp{5cm}} 
    \hline\hline
    \textbf{Log Field}         & \textbf{\hspace{2cm}Sample}    \\ 
    \hline
    Time                       & \hspace{2cm} 1         \\
    Source User@domain         & \hspace{2cm} U32@DOM1  \\
    Destination User@domain    & \hspace{2cm} U32@DOM1  \\
    Source Computer            & \hspace{2cm} C815      \\
    Destination Computer       & \hspace{2cm} C625      \\
    Authentication Type        & \hspace{2cm} Kerberos  \\
    Logon Type                 & \hspace{2cm} Network   \\
    Authentication Orientation & \hspace{2cm} LogOn     \\
    Success/Failure            & \hspace{2cm} Success   \\
    \hline\hline
    \end{tabular}}
    \label{tab: authentication log overall}
\end{table}

\subsection{Authentication Log Analysis}
% 身份验证日志是记录内部网络中用户或系统在访问或尝试访问计算机系统时所发生的验证事件的日志文件。它们通常包括成功和失败的登录尝试、时间戳、用户或系统的身份标识符、验证类型（如密码、双因素验证等）、来源IP地址等详细信息。通过分析来自洛斯阿拉莫斯国家实验室（LANL）的真实计算机网络数据集，我们总结了表1所示的日志字段信息。以一条日志数据为例，它记录了时刻（time: 1）发生的一次身份验证事件，在该事件中，用户（U32@DOM1）从源计算机（C815）向目标计算机（C625）发起身份验证请求。请求由用户发起并针对同一用户进行验证。认证类型使用了Kerberos协议，登录类型为网络登录（Network），认证方向是登录（LogOn），且操作成功（Success）。通过这些信息，我们可以追踪网络中的身份验证活动，识别潜在的异常行为。
Authentication logs record authentication events occurring when a user or system accesses or attempts to access a computer system within an internal network. These logs typically include detailed information such as successful and failed login attempts, timestamps, user or system identifiers, authentication types (e.g., password, two-factor authentication), source IP addresses, and more. By analyzing real-world computer network datasets from Los Alamos National Laboratory (LANL)~\cite{kent2016cyber}, we summarize the log field information as shown in Table~\ref{tab: authentication log overall}. 
\begin{figure}[!htb] 
    \centering 
    \includegraphics[width=0.8\linewidth]{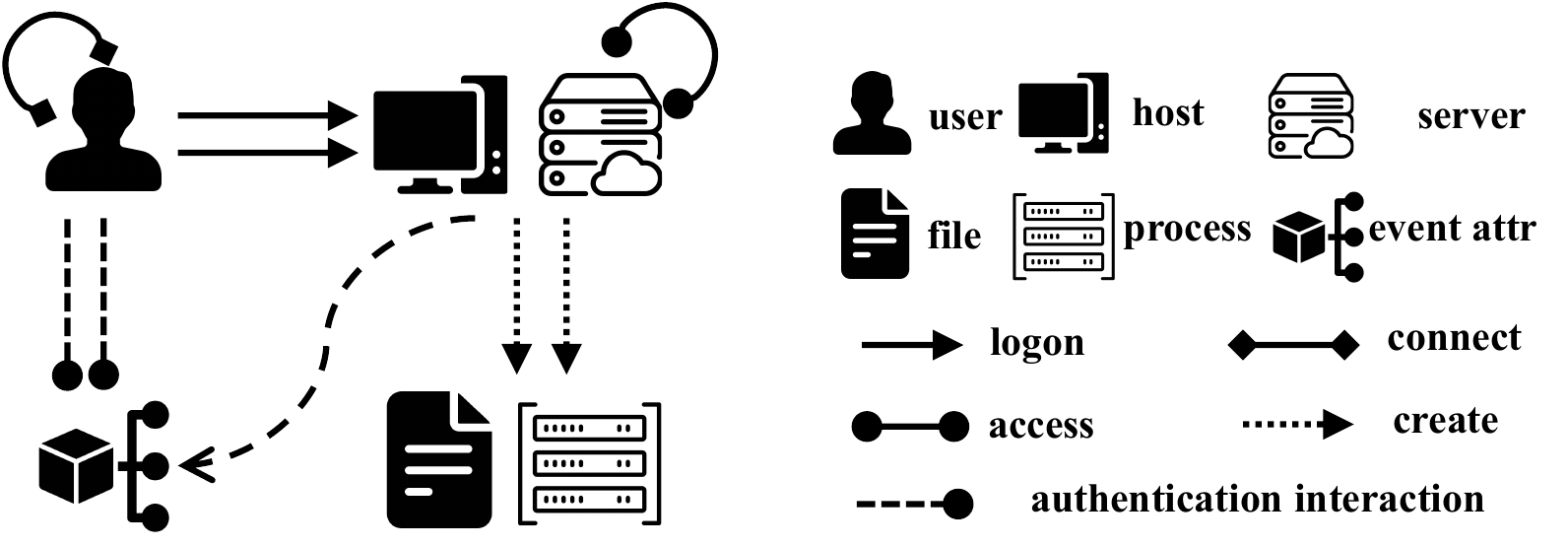} 
    \caption{Illustration of the Heterogeneous Authentication Multigraph.} 
\label{fig: HAG}
\end{figure}
For example, one log entry records an authentication event that occurred at time 1, in which a user (U32@DOM1) initiated an authentication request from a source computer (C815) to a target computer (C625). The authentication type is the Kerberos protocol, the login type is Network login, the authentication direction is LogOn, and the operation is successful. Through this log information, we can track authentication activities within the network and identify potential abnormal behaviors.

% 进一步，我们根据身份验证日志的性质，总结了一些关键结论：
% ① 网络事件核心组成（Core components of network events）：在一个基本的内部网络空间中，用户、主机或服务器设备等实体是不可或缺的，它们的交互构成了整个内部网络生态系统；
% ② 网络事件扩展组成（extension compositions of network events）：进程、文件等实体是用户、主机设备、服务器交互所衍生的网络产物，通过将这些元素纳入分析，我们可以更全面地刻画各种网络行为，从而提高对认证事件的理解；
% ③ 网络事件属性组成（attribute components of network events）：认证类型、登录类型、认证方向分别描述了验证用户身份的协议或方法、用户登录会话的性质、用户请求登录或者注销，这些信息有利于我们对网络行为进行更精确的建模；
% ④ 复杂交互揭示（Complex interactions reveal）：不同网络实体之间存在复杂且多样化的交互，捕捉更加细粒度的交互特征可以实现对网络活动更精细的建模。
% 基于上述结论，我们尝试将大量的身份验证日志构建为一个异构认证图，如图2所示。
Furthermore, based on the characteristics of the authentication logs, we summarize several key conclusions for processing log data:
\begin{itemize}[leftmargin=10pt]
    \item \textbf{Core components of network events:} In a fundamental internal network environment, entities such as users, hosts, or server devices are indispensable, and their interactions form the ecosystem of the internal network.
    \item \textbf{Extension compositions of network events:} Entities like processes and files are network artifacts derived from the interactions between users, hosts and servers. By incorporating these entities, we can more comprehensively characterize various network behaviors, thereby enhancing our understanding of authentication events.
    \item \textbf{Attribute components of network events:} Authentication type, login type, and authentication direction describe the protocol used for user identity verification, the nature of the user login session, and whether the user is requesting login or logout, respectively. The information facilitates more precise modeling of network behavior.
    \item \textbf{Interaction components of network events:} There are diverse interactions between different network entities. Capturing more fine-grained interaction features allows for more detailed modeling of network activities.
\end{itemize}
\subsection{Graph Construction}
Based on these conclusions, we construct a heterogeneous authentication multigraph (\graph) from a large volume of authentication logs, as illustrated in Figure~\ref{fig: HAG}.
% 具体而言，我们首先对每一条身份认证日志条目进行了解析，提取关键字段，如时间戳、用户ID、主机ID、验证事件类型等，并对这些字段进行了相应统一格式化和标准化处理。我们归纳了6种不同的实体（用户、主机、服务器、文件、进程、事件属性）以及5种不同的交互关系（登录、连接、访问、创建、认证）。我们分别将实体和关系作为HAG的节点和边，得到如下的异构认证图。
Specifically, we first parse each authentication log entry to extract key fields such as time, user ID, host ID and authentication event type, and standardize and format these fields uniformly. We then categorize five different types of entities (user, host, server, file, process) as well as four different types of interaction relationships (login, connection, access, creation). Considering that not all of the aforementioned entities are necessarily involved in a single authentication log entry, we further group these five types of entities into three broad categories: \textbf{U}ser-class (U: user), \textbf{D}evice-class (D: host, server) and \textbf{O}bject-class (O: file, process).
These entities and relationships are represented as nodes and edges of \graph, resulting in the following definition:
\begin{definition}
    \emph{\textbf{Heterogeneous Authentication Multigraph (\graph)}}.
    A heterogeneous authentication multigraph can be denoted as $\mathcal{G}=\{\mathcal{V},\mathcal{E},\boldsymbol{X},\boldsymbol{E}, \mathcal{T}^\mathcal{V}, \mathcal{T}^\mathcal{E}, \mathcal{Y}, \mathcal{Z}\}$, where $\mathcal{V}=\{v_1,v_2,\cdots,v_\textit{n}\}$ is the set of nodes (entities), each with a specific type indicated by $\mathcal{T}^\mathcal{V}: v \mapsto$ \{\textit{user}, \textit{host}, \textit{server}, \textit{file}, \textit{process}\}, $\mathcal{E}=\left\{e_{\textit{i},\textit{j}}^\textit{t} \mid e_{\textit{i},\textit{j}}^\textit{t}=\left(v_\textit{i}, v_\textit{j}, t \right), v_\textit{i}, v_\textit{j} \in \mathcal{V}\right\}$ is the set of edges (interaction relations), each with a specific type indicated by $\mathcal{T}^\mathcal{E}: e \mapsto$ \{\textit{login}, \textit{connection}, \textit{access}, \textit{creation}\}, along with a timestamp $t$. $\boldsymbol{X}=\left[\boldsymbol{x}_{1}, \boldsymbol{x}_{2}, \cdots, \boldsymbol{x}_\textit{n}\right]^\top\in\mathbb{R}^{|\mathcal{V}|\times d_\textit{v}}$ is the node feature matrix, where $\boldsymbol{x}_\textit{i}$ is the feature vector constructed by concatenating the type information (one-hot encoded) of node $v_\textit{i}$ and its ID information, and $d_\textit{v}$ is the dimension of node features.
    $\boldsymbol{E}=\left[\boldsymbol{e}_1,\boldsymbol{e}_2,\cdots,\boldsymbol{e}_\textit{m}\right]^\top\in\mathbb{R}^{|\mathcal{E}|\times d_\textit{e}}$ is the edge feature matrix, where $\boldsymbol{e}$ is the edge feature vector constructed by concatenating the information of interaction type, authentication type, logon type and authentication orientation, and $d_\textit{e}$ is the dimension of edge features.
    The authentication event $z\in\mathcal{Z}$ involves interactions between three broad categories of entities, which can be represented as a quintuple $z=\langle t,U,D,O\rangle $, and its label information $y_\textit{i}$ is included in $\mathcal{Y}=\{(z_\textit{i},y_\textit{i})\mid z_\textit{i}\in \mathcal{Z}\}$.
\end{definition}
% 注意多图意味着实体之间可能存在多重交互，通过不同时间戳来区分。由于实体类型和关系类型的数量是已知的，因此特征的维度是可以确定的，即d=7. 最后，HAMG可以有效地表示复杂的网络活动，为后续的横向移动检测提供强大的数据结构基础
Note that a multigraph implies that multiple interactions may exist between entities, which are distinguished by different timestamps. 
% Since the number of entity types and relationship types is known, the feature dimensions can be determined, i.e.,  $d_\textit{v}=6+1=7$, $d_\textit{e}=5$. 
Finally, \graph can effectively represent complex network activities, providing a powerful data structure for subsequent lateral movement detection.

\section{Methodology}
% In this section, we will introduce the detection process of \textit{LMDetect} in detail. The framework of \textit{LMDetect} is shown in Fig~\ref{fig: LMDetect}. \textit{LMDetect} constructs a Heterogeneous Authentication Graph (HAG) and takes as input the authentication event tuples $z_i$, and the output is the labeling of the authentication event behavior $y_i$.
% Our detection framework mainly consists of the following components: (a) Heterogeneous Authentication Graph Construction: constructing a Heterogeneous Authentication Graph (HAG) and Temporal Interaction Mapping $TempIntMap$ for authentication logs; (b) Time-aware Subgraph Generator (TSG): this component generates time-aware subgraphs centered on authentication events based on temporal information; (c) Multi-scale Attention Encoder and Prediction: this module characterizes the subgraphs through a multi-scale attention mechanism, and takes the learned graph characterizations and makes classification predictions. HAG construction has been described in the previous section, next, we will describe the functionality and implementation of the other components in detail.

\begin{figure*}[!htb] 
    \centering 
    \includegraphics[width=1\textwidth]{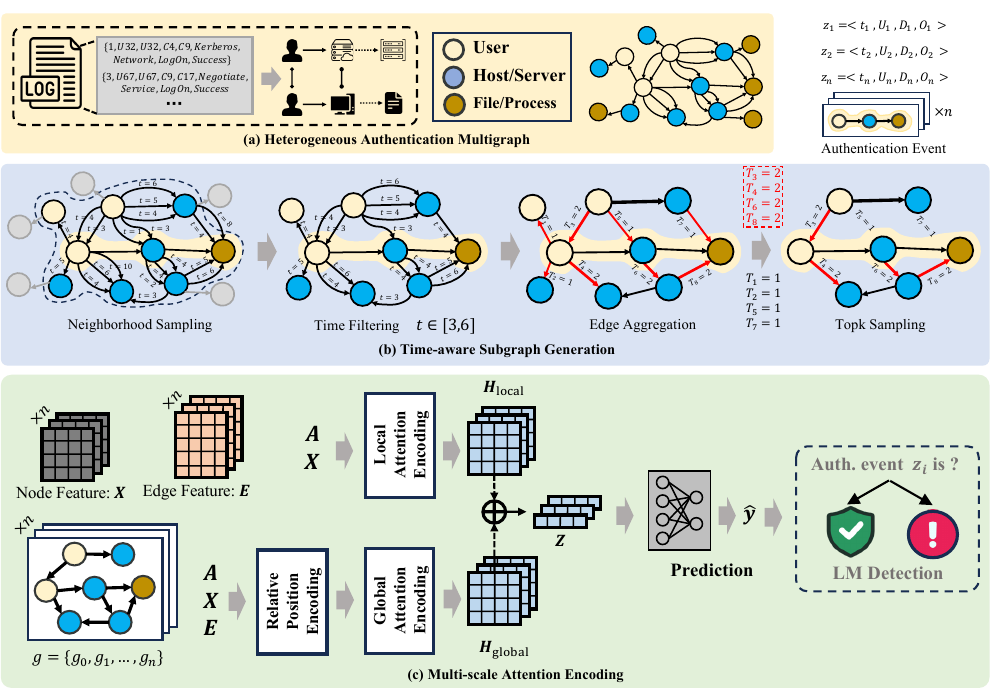} 
    \caption{Illustration of \textit{LMDetect} framework. The complete workflow is as follows: 1) Constructing heterogeneous authentication multigraph using authentication logs; 2) Sampling authentication subgraphs for target events via time-aware subgraph generator; 3) Learning the behavior patterns of authentication events via multi-scale attention encoder and detecting lateral movement behavior via subgraph classification.} 
\label{fig: LMDetect}
\end{figure*}

% \subsection{Problem Definition}
% 由数以亿计的身份认证日志转化而来的HAMG规模十分庞大，将会极大地增加后续图分析与检测的计算负担。此外，现有的研究往往将横向移动检测建模为边分类或路径分类问题，忽略了网络行为的持续性和多阶段问题。考虑到现有方法在检测范式和计算效率上的局限性，本工作创新地将横向移动检测建模为子图分类问题，其理由如下：1）横向移动行为往往涉及大量实体间的一系列交互，其结构可以被横向移动路径引导的子图所覆盖；2）包含横向移动行为的子图在规模上远小于整个HAMG，可以结合小批量训练方式来代替高计算复杂度的全批量训练；3）基于子图的归纳式学习可以提高模型对未见行为的识别能力，使其更好地在不平衡的数据环境下适应变化的横向移动行为模式。

% 基于此，我们设计了一个基于动态感知子图分类的横向移动检测框架，由两个模块构成：1）动态感知子图生成器，用于采样目标认证事件周围的高相关信息，形成以目标认证事件为中心的时间感知认证子图；2）多尺度注意力编码器，用于学习网络实体间的局部和全局依赖，实现对认证事件的强大表征。该框架的示意图如图所示

The \graph derived from billions of authentication logs is extraordinarily large, imposing a significant computational burden for subsequent analysis and detection. Additionally, existing studies often model lateral movement detection as either edge (path) classification or link prediction problem, neglecting the continuity and multi-stage nature of network behavior. Given the limitations of existing approaches in terms of detection paradigms and computational efficiency, we innovatively models lateral movement detection as a \textbf{subgraph classification} problem for the following reasons: 1) Lateral movement behaviors often involve a series of interactions among numerous entities, whose structure can be captured by subgraphs induced from lateral movement paths; 2) Subgraphs containing lateral movement behaviors are much smaller in size compared to the entire \graph, enabling mini-batch training to replace computationally expensive full-batch training; 3) Inductive learning based on subgraphs enhances the model's ability to identify unseen behaviors, allowing it to better adapt to evolving lateral movement patterns in imbalanced data environments.

Based on this, we design a time-aware subgraph classification framework named \model for lateral movement detection, comprising two main modules: 1) a time-aware subgraph generator, which samples highly relevant information surrounding the target authentication event to form a time-aware authentication subgraph centered on the target event; 2) a multi-scale attention encoder, which learns both local and global dependencies among network entities, providing powerful representations of authentication events. The schematic of this framework is illustrated in Figure\ref{fig: LMDetect}.

% TODO

% TODO: 子图生成的伪代码
\begin{algorithm}[!htb]
    \caption{Time-aware Subgraph Generation}
    \label{alg: sample}
    \begin{algorithmic}[1]
        \REQUIRE {Target authentication event $z_\textit{i}=\langle t_\textit{i},U_\textit{i}, D_\textit{i}, O_\textit{i}\rangle $, \graph, time window parameter $\tau$, topk parameter $k$.}
        \ENSURE {Time-aware subgraph $g_\textit{i}$.}
            \STATE Get 1-hop neighbors of the involved entities in event $z_\textit{i}$ and merge them:  
            $\mathcal{N}_{{z}_\textit{i}} = \mathcal{N}_{U_\textit{i}} \cup \mathcal{N}_{D_\textit{i}} \cup \mathcal{N}_{O_\textit{i}}$;
            \STATE Extract subgraph $g_\textit{i} = \left\{\mathcal{V}_{{g}_\textit{i}}, \mathcal{E}_{{g}_\textit{i}}\right\}$ from \graph containing all nodes in $\mathcal{N}_{{z}_\textit{i}}$ along with all the edges between them;
            \STATE Remove edges from $g_\textit{i}$ that fall outside $[t_\textit{i}-\tau, t_\textit{i}+\tau]$: \\$\mathcal{E}_\textit{g} \leftarrow \mathcal{E}_\textit{g} \setminus \left\{e_\textit{k}\mid e_\textit{k}=\left(v_\textit{i}, v_\textit{j}, t \right), v_\textit{i}, v_\textit{j} \in \mathcal{V}_\textit{g}, |t-t_\textit{i}|>\tau\right\}$;
            \STATE Merge multiple edges between pairwise nodes and generate edge feature $T$ represented interaction frequency;
            \STATE Mark $U_\textit{i}, D_\textit{i}, O_\textit{i}$ as core nodes and other nodes in $\mathcal{N}_{{z}_\textit{i}}$ as auxiliary nodes;
            \STATE Sort the edges between core nodes and auxiliary nodes in descending order based on $T$ and retain the top-$k$ edges with the highest $T$, yielding the time-aware subgraph $g_\textit{i}$;
            \STATE Assign the label $y_\textit{i}$ of event $z_\textit{i}$ to $g_\textit{i}$;
        \RETURN {Time-aware subgraph $g_\textit{i}$.}
    \end{algorithmic} 
\end{algorithm}

\subsection{Time-aware Subgraph Generation}\label{Section TSG}
% 重新回顾身份认证事件的定义，即$z=\langle t,U,D,O\rangle $，其中t表示事件发生的时间，$U,D,O$表示事件所涉及的3类实体。我们的时间感知子图生成器将会接受事件作为输入，从\graph中采样对应的时间感知子图作为输出。具体而言，对于事件z涉及的三个实体，我们将它们视为核心节点，分别获取它们的一阶邻居集合并进行合并：
Revisiting the definition of an authentication event $z_\textit{i}=\langle t_\textit{i},U_\textit{i}, D_\textit{i}, O_\textit{i}\rangle $, where $t_\textit{i}$ denotes the time of the event, and $U_\textit{i}, D_\textit{i}, O_\textit{i}$ represent the three types of entities involved in the event. Our time-aware subgraph generator takes the event as input and samples the corresponding time-aware subgraph from the graph as output. Specifically, for the three entities involved in event $z$, we treat them as core nodes and retrieve their 1-hop neighbor sets, which are then merged:
\begin{equation}
    \mathcal{N}_{{z}_\textit{i}} = \mathcal{N}_{U_\textit{i}} \cup \mathcal{N}_{D_\textit{i}} \cup \mathcal{N}_{O_\textit{i}}
\end{equation}
% 注意三个核心节点也包含在该集合中，即$U_\textit{i}, D_\textit{i}, O_\textit{i}\in \mathcal{N}_\textit{z}$。我们将该集合中的其他节点视为辅助节点。然后我们从\graph中导出包含\mathcal{N}_\textit{z}中所有节点以及相关连边的子图$g=\left\{\mathcal{V}_\textit{g}, \mathcal{E}_\textit{g} \right\}$ (为了简化表示，我们对子图的符号定义仅保留了节点集和边集)。
% 我们以事件z发生的时间为中心设定了一个时间区间，并从子图中移除那些不在该时间区间内发生的交互（边），得到新的边集合。
Note that the three core nodes are also included in this set, \ie $U_\textit{i}, D_\textit{i}, O_\textit{i}\in \mathcal{N}_{{z}_\textit{i}}$. We treat other nodes in this set as auxiliary nodes. Subsequently, we extract the subgraph $g_\textit{i} = \left\{\mathcal{V}_{{g}_\textit{i}}, \mathcal{E}_{{g}_\textit{i}}\right\}$ from the graph, containing all nodes in $\mathcal{N}_{{z}_\textit{i}}$ along with their corresponding edges (For simplicity, the notation for the subgraph retains only the node set and edge set).
Centered on the time $t_\textit{i}$ when event $z_\textit{i}$ occurs, we define a time interval and remove interactions (edges) from the subgraph that fall outside this interval:
\begin{equation}
    % \small
    \mathcal{E}_{\textit{g}_\textit{i}} \leftarrow \mathcal{E}_{\textit{g}_\textit{i}} \setminus \left\{e_{\textit{i},\textit{j}}^\textit{t}\mid e_{\textit{i},\textit{j}}^\textit{t}=\left(v_\textit{i}, v_\textit{j}, t \right), v_\textit{i}, v_\textit{j} \in \mathcal{V}_{\textit{g}_\textit{i}}, |t-t_\textit{i}|>\tau\right\} 
\end{equation}
% 紧接着，考虑到子图存在多边的情况，我们首先对子图进行多边合并，即对同一对节点之间存在的多次交互进行合并，并新增一个表示交互次数(T)的边特征，形成新的压缩子图。然后，我们对压缩子图中核心节点和辅助节点之间的边按照交互次数进行降序排列，并保留交互次数topk的边，最终得到时间感知的子图。
Considering the existence of multiple edges in the subgraph, we first perform edge aggregation, merging multiple interactions between the same pair of nodes. 
A new edge feature $T$ is introduced to represent the interaction frequency, forming a compressed subgraph. Subsequently, we sort the edges between core nodes and auxiliary nodes in descending order based on the interaction frequency and retain the top-$k$ edges with the highest interaction frequency, resulting in the final time-aware subgraph.
For each authentication event $z\in\mathcal{Z}=\left\{z_1,z_2,\cdots,z_\textit{n}\right\}$, we sample the corresponding time-aware subgraph $g_\textit{i}$ from \graph and assign the label of the authentication event $z_\textit{i}$ to it, ultimately forming a subgraph dataset $\mathcal{D}=\left\{\left(z_\textit{i}, g_\textit{i}, y_\textit{i}\right) \mid z_\textit{i}\in\mathcal{Z} \right\}$.

%

% 图4例证了整个时间感知子图生成的过程。注意，邻域采样保留了直接的交互实体，时间过滤筛选短时间内的最相关的交互行为，而边聚合与基于交互频率的topk采样进一步捕捉了时间感知的重要子结构。该时间感知子图有效地保留了认证事件所包含的关键信息，为后续的横向移动检测提供了高质量的数据基础。
Figure~\ref{fig: LMDetect}(b) and Algorithm~\ref{alg: sample} illustrates the entire process of time-aware subgraph generation. Notably, neighborhood sampling retains directly interacting entities, while time filtering selects the most relevant interactions within a short time window. Edge aggregation and top-$k$ sampling based on interaction frequency further capture important time-aware substructures. This time-aware subgraph effectively preserves the key information contained in the authentication event, providing a high-quality data foundation for subsequent lateral movement detection.

\subsection{Multi-Scale Attention Encoding}
% 得到认证事件的子图数据集后，我们进一步设计了一个多尺度注意力编码模块来有效地捕捉认证事件的行为模式，检测潜在的恶意行为。具体而言，多尺度注意力编码模块由3部分组成：1）局部注意力编码器刻画认证子图中实体间的局部交互，捕捉细粒度地认证行为模式；2）相对位置编码器模拟认证行为的移动路径，捕捉认证子图的拓扑结构信息；3）全局注意力编码器刻画认证子图中实体间的全局交互，捕捉长距离的行为依赖。图3展示了多尺度注意力编码模块的完整架构。
After obtaining the subgraph dataset of authentication events, we further design a multi-scale attention encoding module to effectively capture behavior patterns of authentication events and detect potential malicious actions. Specifically, the multi-scale attention encoding module consists of three components: 1) a local attention encoder that characterizes local interactions between entities in the authentication subgraph, capturing fine-grained authentication behavior patterns; 2) a relative positional encoder that models the movement paths of authentication behaviors, capturing the topological structure information of the authentication subgraph; 3) a global attention encoder that characterizes global interactions between entities in the authentication subgraph, capturing long-distance behavioral dependencies. Figure~\ref{fig: LMDetect}(c) illustrates the architecture of the multi-scale attention encoding module.

\subsubsection{Local Attention Encoding}
% 虽然时间感知子图从时间尺度上覆盖了最具动态相关性的实体和交互，但不同交互对象对目标认证事件的表征具有不同的贡献度。例如，一个包含用户、主机和服务器的内部网络，如果目标身份验证事件涉及用户$u_1$在主机$h_1$上的登录操作，那么主机$h_1$的身份验证日志和服务器$s_1$的通信记录可能提供与目标事件更直接相关的上下文信息。相比之下，其他主机$h_2$或$h_3$在时间窗口内与用户$u_1$通信的记录虽然也相关，但对目标身份验证事件的贡献程度相对较低。为了有效地量化这种贡献差异，我们引入了局部注意力计算，来学习认证子图中任意实体间的相互贡献权重。
Although the time-aware subgraph covers entities and interactions with the most temporally relevance, different interacting entities contribute variably to the representation of the target authentication event. For instance, in an internal network containing users, hosts, and servers, if the target authentication event involves user $u_1$ logging in to host $h_1$, then the authentication logs of host $h_1$ and the communication records with server $s_1$ may provide more directly relevant contextual information for the target event. In contrast, logs of other hosts $h_2$ or $h_3$ interacting with user $u_1$ within the time window, while still relevant, contribute comparatively less to the target authentication event. To effectively quantify these contribution differences, we introduce local attention computation to learn the mutual contribution weights between any entities within the authentication subgraph.
Specifically, for a target node $v_\textit{i}$ in subgraph $g$, the attention weight (contribution) of its neighbors $v_\textit{j}\in\mathcal{N}_{v_\textit{i}}$ can be represented as follows:
\begin{equation} \label{eq: node-att}
    \footnotesize
    \alpha_{\textit{i},\textit{j}}^{(l)}
    =\frac{\exp\left(\text{LeakyReLU}\left(\boldsymbol{a}^\top\left[\boldsymbol{\Theta}_\textit{n}^{(l)}\boldsymbol{h}_\textit{i}^{(l)}\parallel\boldsymbol{\Theta}_\textit{n}^{(l)}\boldsymbol{h}_\textit{j}^{(l)}\right]\right)\right)}{\sum_{o\in\mathcal{N}_{v_\textit{i}}}\exp\left(\text{LeakyReLU}\left(\boldsymbol{a}^\top\left[\boldsymbol{\Theta}_\textit{n}^{(l)}\boldsymbol{h}_\textit{i}^{(l)}\parallel\boldsymbol{\Theta}_\textit{n}^{(l)}\boldsymbol{h}_\textit{o}^{(l)}\right]\right)\right)}
\end{equation}
where $\boldsymbol{h}^{(l)}$ is the node feature vector and $\boldsymbol{h}^{(0)}=\boldsymbol{x}$, $\boldsymbol{\Theta}$ is the weight matrix used to transform node features at layer $l$, $\boldsymbol{a}$ is the attention parameter vector for calculating similarity between nodes, LeakyReLU is the activation function, and $\parallel$ denotes the concatenation operation.
Once the attention scores are obtained, they are used to update the target node's features by aggregating information from its neighborhood:
\begin{equation} \label{eq: local-att-encoder} 
    \boldsymbol{h}_{\textit{i}, \text{local}}^{(l+1)}=\text{Elu}\left(\sum_{v_\textit{j}\in\mathcal{N}_{v_\textit{i}}\cup\{v_\textit{i}\}}\alpha_{\textit{i},\textit{j}}^{(l)}\boldsymbol{\Theta}_{\alpha}^{(l)}\boldsymbol{h}_\textit{j}^{(l)}\right)
\end{equation}
where a transformation parameterized by weights $\boldsymbol{\Theta}_{\alpha}^{(l)}$ and Elu activation is used to generate the final node features. The local attention mechanism enables us to adaptively assign different weights to neighboring nodes, thereby capturing their different contributions to the target nodes.
We represent the node feature matrix generated via local attention encoding as $\boldsymbol{H}_{\text{local}}^{(L)}$, where $L$ is the number of layers for local attention encoding.

\subsubsection{Relative Position Encoding}
% 网络身份认证活动涉及大量实体间的复杂连续交互，而这其中涉及的横向移动行为可以被视为一种在网络实体间不断跳跃的有向交互。为了有效捕捉认证子图的结构信息，我们引入了相对位置编码模块，为后续全局特征编码提供结构信息。对于每个认证子图$g$，相对位置编码器通过多阶随机游走来建模实体间的直接和间接交互，捕捉实体间的相对位置关系，并增广特征空间。具体而言，图上的随机游走可以理解为一个马尔可夫过程。在随机游走过程中，从某个节点$v_\textit{i}$出发，随机选择一条相连的边，到达其邻居节点。这一过程的选择是由转移概率决定的。在简单随机游走中，转移概率通常设定为$\boldsymbol{M}_\textit{ij}=\frac{\boldsymbol{A}_\textit{ij}}{\boldsymbol{D}_\textit{ii}}$，表示从节点$v_\textit{i}$出发，有$\boldsymbol{D}_\textit{ii}$个邻居节点，每个节点被选择的概率是均等的。进一步，转移矩阵可以表示为$\boldsymbol{M}=\boldsymbol{D}_\textit{g}^{-1}\boldsymbol{A}_\textit{g}$，其中$\boldsymbol{A}_\textit{g}$和$\boldsymbol{D}_\textit{g}$分别表示子图$g$的邻接矩阵和度矩阵。
%
% 为有效捕捉认证子图中的结构信息，我们引入了相对位置编码模块，以便在全局特征编码过程中提供结构性支持。具体来说，对于每个认证子图 $g$，相对位置编码器通过多阶随机游走来建模实体间的直接和间接交互，从而捕捉实体之间的相对位置关系，进而丰富特征空间。图上的随机游走可以视为一个马尔可夫过程。在此过程中，随机游走从某一节点 $v_i$ 开始，随机选择一条相连的边，并到达该边的邻居节点。转移的选择由转移概率控制。在简单随机游走中，转移概率通常定义为 $\boldsymbol{M}{ij}=\frac{\boldsymbol{A}{ij}}{\boldsymbol{D}{ii}}$，其中节点 $v_i$ 有 $\boldsymbol{D}{ii}$ 个邻居节点，每个邻居节点被选择的概率是相等的。进一步地，转移矩阵 $\boldsymbol{M}$ 可以表示为 $\boldsymbol{M} = \boldsymbol{D}_g^{-1}\boldsymbol{A}_g$，其中 $\boldsymbol{A}_g$ 和 $\boldsymbol{D}_g$ 分别表示子图 $g$ 的邻接矩阵和度矩阵。
Network authentication activities involve complex, continuous interactions among numerous entities, with lateral movement behavior often manifesting as directed interactions traversing network entities. To effectively capture structural information within the authentication subgraph, we introduce a relative position encoding module to provide structural information during global feature encoding. Specifically, for each authentication subgraph $g$, the relative position encoder models both direct and indirect interactions between entities through multi-step random walks, thereby capturing relative position relationships between entities and enriching the feature space. A random walk on the graph can be regarded as a Markov process. In this process, the random walk starts from a node $v_\textit{i}$, randomly selecting a connected edge to reach a neighboring node. The transition choice is controlled by transition probabilities. In a simple random walk, the transition probability is typically defined as $\boldsymbol{M}_{\textit{i},\textit{j}}=\frac{\boldsymbol{A}_{\textit{i},\textit{j}}}{\boldsymbol{D}_{\textit{i},\textit{i}}}$, where node $v_\textit{i}$ has $\boldsymbol{D}_{\textit{i},\textit{i}}$ neighboring nodes, each with an equal probability of being selected. Furthermore, the transition matrix $\boldsymbol{M}$ can be expressed as $\boldsymbol{M}=\boldsymbol{D}_\textit{g}^{-1}\boldsymbol{A}_\textit{g}$, where $\boldsymbol{A}_\textit{g}$ and $\boldsymbol{D}_\textit{g}$ denote the adjacency matrix and degree matrix of the subgraph $g$, respectively.
% 基于上述定义，$K$步随机游走的转移概率矩阵$\boldsymbol{M}^K$表示在$K$步后从一个节点转移到另一个节点的概率分布，可以通过矩阵$\boldsymbol{M}$的$K$次幂得到：
Based on the above definition, the transition probability matrix of a $K$-step random walk, $\boldsymbol{M}^{(K)}$, represents the probability distribution of transitioning from one node to another after $K$ steps and can be obtained by raising the matrix $\boldsymbol{M}$ to the $K$-th power:
\begin{equation}
    \boldsymbol{M}^{(K)} = \underbrace{\boldsymbol{M}\times\boldsymbol{M}\times \cdots \times \boldsymbol{M}}_K = \boldsymbol{M}^K
\end{equation}
% 相对位置编码器通过计算一至多步转移概率来建模认证子图中实体之间的相对位置：
The relative position encoder models the relative positions between entities in the authentication subgraph by calculating the transition probabilities for one to multiple steps:
\begin{equation}
    \boldsymbol{P}_{\textit{i},\textit{j}}=\left[\boldsymbol{I}, \boldsymbol{M}, \boldsymbol{M}^2,\cdots,\boldsymbol{M}^{K-1}\right]_{i,j} \in\mathbb{R}^{K}
\end{equation}
% 相对位置特征将被进一步整合到节点和边特征中：
Furthermore, the relative position features will be integrated into the node and edge features:
\begin{equation}
    \hat{\boldsymbol{X}}=
    \begin{bmatrix}
        \hat{\boldsymbol{x}}_1
        \\\vdots
        \\\hat{\boldsymbol{x}}_\textit{i}
        \\\vdots
        \\\hat{\boldsymbol{x}}_\textit{n}
    \end{bmatrix}
    =
    \begin{bmatrix}
        \boldsymbol{x}_{1} \parallel \boldsymbol{P}_{1,1}
        \\\vdots
        \\\boldsymbol{x}_\textit{i} \parallel \boldsymbol{P}_{\textit{i},\textit{i}}
        \\\vdots
        \\\boldsymbol{x}_\textit{n} \parallel \boldsymbol{P}_{\textit{n},\textit{n}}
    \end{bmatrix}
    \in\mathbb{R}^{|\mathcal{V}_\textit{g}|\times(d_\textit{v}+K)}
\end{equation}
\begin{equation}
    % \small
    \hat{\boldsymbol{E}}=
    \begin{bmatrix}
        \hat{\boldsymbol{e}}_{1,1}
        \\\hat{\boldsymbol{e}}_{1,2}
        \\\vdots
        \\\hat{\boldsymbol{e}}_{\textit{i},\textit{j}}
        \\\vdots
        \\\hat{\boldsymbol{e}}_{\textit{n},\textit{n}-1}
        \\\hat{\boldsymbol{e}}_{\textit{n},\textit{n}}
        % \\\hat{\boldsymbol{e}}_\textit{n}
    \end{bmatrix}
    =
    \begin{bmatrix}
        \boldsymbol{e}_{1,1} \parallel \boldsymbol{P}_{1,1}
        \\\boldsymbol{e}_{1,2} \parallel \boldsymbol{P}_{1,2}
        \\\vdots
        \\\boldsymbol{e}_{\textit{i},\textit{j}} \parallel \boldsymbol{P}_{\textit{i},\textit{j}}
        \\\vdots
        \\\boldsymbol{e}_{\textit{n},\textit{n}-1} \parallel \boldsymbol{P}_{\textit{n},\textit{n}-1}
        \\\boldsymbol{e}_{\textit{n},\textit{n}} \parallel \boldsymbol{P}_{\textit{n},\textit{n}}
    \end{bmatrix}
    \in\mathbb{R}^{|\mathcal{V}_\textit{g}|^2\times(d_\textit{e}+1+K)}
\end{equation}
We set $\boldsymbol{e}_{\textit{i},\textit{j}}=\boldsymbol{0}$ if $e_{\textit{i},\textit{j}}\notin\mathcal{E}_\textit{g}$.
In this manner, we are able to provide the model with global topological information, thereby facilitating the capture of long-range dependencies among network entities and augmenting the model's capacity to perceive the network structure.

\begin{figure}[!htb] 
    \centering 
    \includegraphics[width=\linewidth]{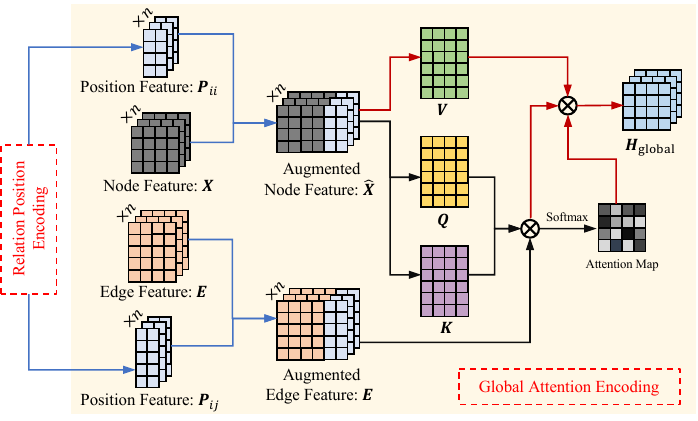} 
    \caption{Illustration of global attention encoding.} 
\label{fig: global-att}
\end{figure}

\subsubsection{Global Attention Encoding}
% 横向移动行为通常表现为攻击者从一个初始节点开始入侵，跨越多个实体，逐步获取更多权限和资源。尽管某些实体之间并不存在直接的连接关系，但通过多跳连接依然可以建立长距离依赖。例如一个用户可以通过多个中间主机来访问特定服务器。因此，仅依靠局部信息编码模块难以全面理解攻击者的行为模式。对此，我们设计了一个基于graph transformer的全局注意力编码器，旨在对融入了相对位置信息的时间感知子图进行全局表征，捕捉网络行为的高阶特征。具体而言，我们的graph transformer利用节点和边特征来计算全局注意力：
Lateral movement behavior typically manifests as an attacker infiltrating the system from an initial node, traversing multiple entities to gradually acquire more privileges and resources. Although certain entities may not be directly connected, long-distance dependencies can still be established through multi-step connections. For example, a user may access a specific server through multiple intermediary hosts. Thus, relying solely on a local information encoding module is insufficient to comprehensively understand the attacker's behavior patterns. To address this, we design a graph transformer as a global attention encoder, as illustrated in Figure~\ref{fig: global-att}, aimed at providing a global representation of the time-aware subgraph enriched with relative position information, thereby capturing higher-order features of network behavior. Specifically, our graph transformer utilizes node and edge features to compute global attention:
\begin{equation} \label{eq: global-att-vector}
        {\boldsymbol{b}}_{\textit{i},\textit{j}}
        =\mathrm{ReLU}\Big(\sigma\left(\left(\boldsymbol{Q}\hat{\boldsymbol{x}}_\textit{i}+\boldsymbol{K}\hat{\boldsymbol{x}}_\textit{j}\right)\odot\boldsymbol{\Theta}_\textit{e}\hat{\boldsymbol{e}}_{\textit{i},\textit{j}}\right)+\boldsymbol{\Theta}_\textit{b}\hat{\boldsymbol{e}}_{\textit{i},\textit{j}}\Big)
\end{equation}
\begin{equation}
    \beta_{\textit{i},\textit{j}}=\frac{\exp{\left(\boldsymbol{\Theta}_\beta{\boldsymbol{b}}_{\textit{i},\textit{j}}\right)}}{\sum_{o\in\mathcal{V}}\exp{\left(\boldsymbol{\Theta}_\beta{\boldsymbol{b}}_{\textit{i},\textit{o}}\right)}}
\end{equation}
where $\boldsymbol{Q}, \boldsymbol{K}, \boldsymbol{\Theta}_\textit{e}, \boldsymbol{\Theta}_\textit{b}, \boldsymbol{\Theta}_\beta$ are the learnable transformation weight matrices, $\odot$ is the Hadamard product, $\sigma(x)=(\text{ReLU}(x))^{1/2}-(\text{ReLU}(-{x}))^{1/2}$ maintains non-linearity and symmetry, which can help alleviate gradient vanishing or exploding issues to some extent.
% 得到全局注意力后，目标节点的特征将通过对所有其他节点以及对应交互的特征进行加权聚合来得到更新
After obtaining the global attention, the target node's features will be updated by performing a weighted aggregation of the features of all other nodes and their corresponding interactions:
\begin{equation} \label{eq: global-feat}
    \hat{\boldsymbol{h}}_{\textit{i},\text{global}}=\sum_{j\in\mathcal{V}}\beta_{\textit{i},\textit{j}}\cdot(\boldsymbol{V}\hat{\boldsymbol{x}}_\textit{j}+\boldsymbol{\Theta}_\textit{g}\boldsymbol{b}_{\textit{i},\textit{j}})
\end{equation}
where $\boldsymbol{V}$ and $\boldsymbol{\Theta}_\textit{g}$ are the learnable transformation weight matrices. The global feature encoding module enables nodes to receive messages from any other node within the subgraph and adaptively integrate external information through global attention. We represent the node feature matrix generated via global attention encoding as $\boldsymbol{H}_\text{global}$.

\subsection{Model Training}
% To achieve lateral movement detection, we use a predictor parameterized by weights $\boldsymbol{\Theta_\textit{p}}$  and a softmax activation to map the subgraph representation into a soft assignment prediction:
% \begin{equation}
%     \hat{\boldsymbol{y}}_\textit{i} = \text{softmax}\left(\boldsymbol{\Theta_\textit{p}}  \boldsymbol{z}_{\textit{g}_\textit{i}} \right)
% \end{equation}

To achieve lateral movement detection, we fuse the local and global features and apply global sum pooling to obtain the final representation of the authentication subgraph:
\begin{equation} \label{eq: final_rep}
    \boldsymbol{z}_{\textit{g}_\textit{i}}=\text{SumPooling}\left(\boldsymbol{H}_{\text{local}}^{(l+1)}+{\boldsymbol{H}}_{\text{global}}\right)
\end{equation}
During model training, we use a prediction head parameterized by weights $\boldsymbol{\Theta_\textit{p}}$ and a softmax activation to detect potential lateral movement behaviors in the target authentication event:
\begin{equation}
    \hat{\boldsymbol{y}}_\textit{i} = \text{softmax}\left(\boldsymbol{\Theta_\textit{p}}  \boldsymbol{z}_{\textit{g}_\textit{i}} \right)
\end{equation}
This prediction head maps the subgraph representation $\boldsymbol{z}_{\textit{g}_\textit{i}}$ to a soft assignment prediction vector $\hat{\boldsymbol{y}}_\textit{i}$, resulting in a classification error with respect to the true labels, which we compute using the cross-entropy loss:
\begin{equation}
    \mathcal{L_{\textit{pre}}} = -\frac{1}{N}\sum_{i=1}^N \boldsymbol{y}_\textit{i} \log\left(\hat{\boldsymbol{y}}_\textit{i}\right)
\end{equation}
where $N$ denotes the total number of event samples, $\boldsymbol{y}_\textit{i}$ denotes the one-hot encoded true label of the $i$-th time-aware subgraph, indicating the presence or absence of lateral movement behavior in the authentication event $z_\textit{i}$.

\begin{table}[!htb] 
    \renewcommand\arraystretch{1.3} 
    \centering                 
    \caption{Dataset Overview.} 
    \label{tab:dataset-overview}                                 
    \resizebox{0.7\linewidth}{!}{
    \begin{tabular}{cccc} 
\hline\hline

\textbf{Dataset} & \textbf{Nodes} & \textbf{Edges}       & \textbf{Duration}  \\ 
\hline
LANL             & 57,816         & 3,914,890       & 30 Days            \\
CERT             & 93,779         & 2,384,122       & 60 Days            \\
\hline\hline
\end{tabular}
}
\end{table}

\section{Experiment}
\subsection{Dataset}
We evaluate \model on two commonly used datasets: LANL~\cite{kent2016cyber} and CERT~\cite{Lindauer2020}. We construct corresponding \graph using these two datasets, with specific information shown in Table~\ref{tab:dataset-overview}.
% We chose two datasets to evaluate our \textit{LMDetect} framework:
\subsubsection{LANL~\cite{kent2016cyber}}
Los Alamos National Laboratory released a comprehensive dataset covering multi-source cybersecurity events in 2015. Spanning 58 days, the dataset includes anonymized event data collected from five sources within the laboratory's internal network, including Windows-based authentication events, system start and stop events on individual machines, DNS queries, network traffic data at router locations, and red team events simulating malicious behavior. The total dataset size is approximately 12 GB, recording over 1,648,275,307 event entries generated by 12,425 users, 17,684 computers, and 62,974 processes. We extract 30 days of event data from this dataset for research and analysis.

% Los Alamos National Lab (LANL) released a comprehensive multi-source cybersecurity incident dataset in 2015. The dataset collects de-identified event data from five sources within the Los Alamos National Lab internal computer network over a 58-day time span. The data sources include Windows-based authentication events from individual computers and centralized Active Directory domain controller servers, processing of start and stop events from individual Windows computers, domain name service (DNS) lookups collected on internal DNS servers, network traffic data collected at several key router locations, and a set of well-defined red team events that mimic malicious behavior over a 58-day period. Overall, the dataset compresses approximately 12 GB of data across five data elements, logging a total of 1,648,275,307 events for 12,425 users, 17,684 computers, and 62,974 processes. We extracted event data from LANL spanning 30 days.

\subsubsection{CERT~\cite{Lindauer2020}}
The Insider Threat Test Dataset from CERT, provided by Carnegie Mellon University, simulates employee activities and potential insider threats in a corporate network. We utilize the latest r6.2 version, which includes over 3.5 million user login events and 2 million file access events. This dataset covers operational activities of 4,000 users and 4,400 computers and includes five types of insider threat scenarios, totaling 470 labeled security events. We extract a two-month continuous sample from the dataset to support cybersecurity behavior analysis.

% The Insider Threat Test Dataset, published by the Software Engineering Institute at Carnegie Mellon University, collects employee behavioral records and attack scenarios from corporate networks, containing 3,530,286 user login events and 2,014,884 file access events, among others.Each event has multiple dimensions of description, e.g., login events include login time, user and computer. The dataset contains 4,000 users and 4,400 computers, of which 400 are shared computers. The dataset also includes five insider threat scenarios with 470 labeled security events. We chose the latest version r6.2 and extracted user behavior records for 2 consecutive months from data spanning 18 months.

% We constructed different HAGs for these two datasets separately and used Table~\ref{tab:dataset-overview}  to show the information of the HAGs obtained from the construction.

\subsection{Comparison Methods}
In order to validate the superiority of \model in lateral movement detection, we compare it with a variety of existing state-of-the-art methods, including GNN-based detection methods as well as several classical graph-based detection systems. Note that traditional detection methods are not included in the comparison because they are no longer able to cope with the current complex attack scenarios.

For GNN-based methods, we select three commonly used GNN models for comparison: GCN~\cite{kipf2016semi}, GAT~\cite{velickovic2017graph} and GraphSAGE~\cite{hamilton2017inductive}. The Graph Convolutional Network (GCN) employs graph convolution operations, combining topological structure to uniformly aggregate information from neighboring nodes into the target node, thereby capturing interaction features among nodes. The Graph Attention Network (GAT) introduces an attention mechanism that adaptively adjusts the target node's reception of information from different neighbors by calculating attention weights between the node and its neighbors. GraphSAGE samples and aggregates the neighborhood information around the target node, enhancing scalability for representation learning in large-scale graphs. We use these three models, following a subgraph classification paradigm, to implement lateral movement detection.
% 注意带有“-ts”后缀的版本表明使用时间感知子图来进行横向移动检测，而无后缀版本表明使用随机采样的子图来进行横向移动检测。
Note that versions with the ``-ts'' suffix indicate the use of time-aware subgraphs for lateral movement detection, whereas versions without the suffix indicate the use of randomly sampled subgraphs.

For graph-based lateral movement detection methods, we select LMTracker~\cite{fang2022lmtracker} and Euler~\cite{king2023euler} for comparison. LMTracker is a static graph-based detection method that models lateral movement detection as a path classification problem, leveraging the metapath2vec~\cite{dong2017metapath2vec} method to learn representations of network entities and generate path information for detection. Euler is a temporal graph-based detection method that uses timestamp information to construct the authentication logs into a temporal graph composed of snapshots. It combines GNNs and recurrent neural networks (RNNs) to capture both topological and temporal features of network activity, enabling lateral movement detection. In this paper, we use combinations of GCN with GRU and LSTM for lateral movement detection, referred to as Euler GCN-GRU and Euler GCN-LSTM, respectively. Note that since the original Euler study did not include experiments on the CERT dataset, we have also excluded Euler's results on the CERT dataset in our work.

% For graph-model based lateral movement detection systems, we chose LMTracker and Euler for comparison. LMTracker is a static graph-based detection system that models and classifies target authentication events as paths in a graph through graph embedding using metapath2vec. In contrast, Euler is a dynamic time-series graph-based detection system, which builds authentication logs into discrete graph snapshots by utilizing timestamp information, and then combines it with a series of GNNs and recurrent neural networks (RNNs) to capture topological and temporal features of the network for link prediction of lateral movement. In this paper, we use a combination of GNN and RNN for GCN-GRU and GCN-LSTM, which we denote by Euler GCN-GRU, Euler GCN-LSTM, respectively. Note that Euler's experimental results are not included in our comparison of the CERT because Euler's researchers did not conduct research on the CERT.

\subsection{Evaluation Metrics}
% We believe that the use of multiple performance metrics can provide a more comprehensive view of the model's performance in real-world scenarios. In previous researches, the performance evaluation of lateral movement detection models is usually limited to one or two metrics. In contrast, whereas we systematically evaluated the \textit{LMDetect} model by five comprehensive metrics: \textbf{precision}, \textbf{recall}, \textbf{F1-score}, \textbf{accuracy}, and Area Under the Curve (\textbf{AUC}).

% In the field of lateral movement detection, evaluation metrics such as overall accuracy may produce misleading results due to the extreme imbalance of positive and negative samples. Therefore, we used \textbf{Macro} metrics to evaluate the detection methods in our experiments. It can more fairly reflect the performance of the model on different categories by calculating the evaluation results for each category separately and averaging the results across all categories. In particular, when dealing with data imbalances, it can more effectively evaluate the performance of a model on small sample categories.

% 在以往的研究中，横向移动检测模型的性能评估通常围绕一到两个指标。而我们采用多个性能指标（precision，recall，F1-score, Accuracy, AUC）来更全面地评估模型在实际应用中的表现。Precision关注正例预测的准确性，适合在误报代价高的场景使用；Recall注重正例覆盖率，有助于减少漏报，尤其在需要识别所有正例的任务中有重要价值；Accuracy则衡量模型的整体预测准确性，但在类别不平衡时可能失去代表性；F1平衡了Precision和Recall，特别适合类别不平衡的场景，提供更全面的正例预测性能评估；AUC则衡量模型在不同阈值下区分正负样本的能力，用于全面评估模型的总体效果。在横向移动检测领域中，由于正负样本分布呈现出极端不平衡，准确率等评估指标可能会产生误导性结果。因此，我们使用宏版本的指标来进行评估。宏版本指标通过分别计算每个类别的评估结果，并对所有类别的结果进行平均处理，可以更公平地反映模型在不同类别上的性能。特别是在处理数据不平衡问题时，宏版本指标能更有效地评估模型在少类别上的表现。

In previous studies, the performance evaluation of lateral movement detection models typically focused on one or two metrics. In contrast, we employ multiple performance metrics (precision, recall, F1-score, accuracy, and AUC) to more comprehensively assess the model's performance in real-world applications. Precision emphasizes the accuracy of positive predictions, making it suitable for scenarios where the cost of false positives is high. Recall focuses on positive coverage, helping reduce false negatives, which is particularly valuable in tasks where identifying all positive instances is crucial. Accuracy measures the overall prediction accuracy of the model but may become less representative in imbalanced class distributions. F1-score balances precision and recall, making it particularly suitable for imbalanced classes, as it provides a more complete evaluation of positive prediction performance. AUC assesses the model's ability to distinguish between positive and negative samples across different thresholds, offering an overall evaluation of the model's effectiveness.

\subsection{Experimental Settings}
The details of our experimental setup are as follows.
The \model is implemented using PyTorch and its associated Geometric library. We use 17,000 benign event samples and 400 malicious event samples on both datasets to simulate the actual sample imbalance problem. With such a sample distribution, we are able to more realistically reproduce the actual network environment, where malicious events occur much less frequently than normal events. This data imbalance property puts higher demands on the model, prompting it to accurately identify a small number of abnormal behaviors even when faced with a large number of normal behaviors. We divide the dataset into training, validation, and testing sets according to the ratio of 6:2:2, and set up 10 independent repeated experiments with different random seeds. 
For different modules in \model, the settings of key parameters are presented in Table~\ref{tab:Experimental Settings}.

\begin{table}
    \renewcommand\arraystretch{1.2}
\centering
\caption{Key parameter settings in \model framework.}
\label{tab:Experimental Settings}
\resizebox{\linewidth}{!}{
    \begin{tabular}{c|c|c|c} 
        \hline\hline
        \multirow{2}{*}{\textbf{Para.}} & \multirow{2}{*}{\textbf{Description}} & \multicolumn{2}{c}{\textbf{Dataset}}  \\ 
        \cline{3-4}
                                            &                                       & \textbf{LANL} & \textbf{CERT}         \\ 
        \hline
        $\tau$              & time window parameter            & 3600s         & 10800s                \\
        % $h$                 &                                & 1             & 1                     \\
        $k$                 & top-$k$ sampling parameter       & 150           & 150                   \\ 
        \hline
        $K$                 & random walk step length            & 32            & 32                    \\
        $L$                 & number of local attention encoding layers          & 2             & 2                     \\
                            & number of global attention heads & 4             & 4                     \\
                            & dropout rate                     & 0.2           & 0.2                   \\
                            & hidden dimensions           & 64            & 64                    \\
                            & batch size           & 16            & 16                    \\
                            & learning rate           & 0.0005            & 0.0005                    \\
        \hline\hline
    \end{tabular}}
\end{table}

% In TSG, $\tau$ denotes the size of the time window. LANL captures 1,648,275,307 authentication events in a time span of 58 days, while the CERT dataset, though large in size, has a time span of 18 months. Therefore, for the LANL dataset, we set the time window size to $\pm$1 hour, whereas for the CERT dataset, we set the time window size to $\pm$3 hours. $h$ is the order of subgraph sampling, and $k$ is the number of sampled neighbors per order of the authentication event entity node.

% In \textit{MSAE}, we set the random walk step size $k-steps$ for relative position encoding to 32. the global attention encoding part uses a multi-head mechanism with $n$ = 4 heads and an attention dropout $p$ of 0.2, and the aggregation method in the local attention encoding part of the GNN is MEAN. in addition, we stacked two \textit{MSAE} encoding layers ($l$ = 2) with a hidden dimension of 64 for embedding against authentication events. In addition, the prediction head $f_{\psi}$ is a feed-forward neural network (FFNN) consisting of a Relu activation function and a linear layer, and we set the batch size, and the learning rate to 16 and 0.0005, respectively.

For GCN, GAT, GraphSAGE, we employ these three GNN models on our constructed \graph. The number of message passing layers is set to 2 for all models, and use the same prediction header for subgraph classification. The batch size, learning rate, data size and data division of these models are all consistent with \model. 
% In addition, we deployed TSG module on these GNN models and named them `GCN-ts', `GAT-ts' and `GraphSAGE-ts', respectively. Our aim is to validate the evaluation of the time-aware subgraph generation module deployed on the base graph model also to achieve the improved detection performance.
For LMTracker and Euler, we conduct the experiments strictly according to the detection process designed in their original papers and follow the parameter settings provided in their papers. The data size and data division are the same as \model. % This strict experimental design ensures the validity and fairness of the results and makes the performance comparison between different methods highly credible.

\begin{table*}[!htb] 
    \renewcommand\arraystretch{1.5} 
    \centering                 
    \caption{Lateral movement detection results: Metric (\%) $\pm$ Standard Deviation. The best results are highlighted in bold, and the second-best results are underlined.} 
    \label{tab:baseline-results}                                 
    \resizebox{\textwidth}{!}{
\begin{tabular}{c|ccccc|ccccc} 
\hline\hline
\multirow{2}{*}{\textbf{Method}} & \multicolumn{5}{c|}{\textbf{LANL}}                                                                                                   & \multicolumn{5}{c}{\textbf{CERT}}                                                                                                     \\ 
\cline{2-11}
                                 & \textbf{Precision}       & \textbf{Recall}          & \textbf{F1}              & \textbf{Accuracy}        & \textbf{AUC}             & \textbf{Precision}       & \textbf{Recall}          & \textbf{F1}              & \textbf{Accuracy}        & \textbf{AUC}              \\ 
\hline
GCN                              & 57.42\tiny{$\pm$0.774}\normalsize          & 63.73\tiny{$\pm$2.076}\normalsize        & 59.24\tiny{$\pm$0.458}\normalsize              & 94.56\tiny{$\pm$0.959}\normalsize       & 76.33\tiny{$\pm$0.736}\normalsize              & 50.70\tiny{$\pm$0.270}\normalsize              & 70.49\tiny{$\pm$2.342}\normalsize                  & 46.03\tiny{$\pm$4.028}\normalsize              & 80.89\tiny{$\pm$11.17}\normalsize                & 81.53\tiny{$\pm$3.339}\normalsize         \\
GAT                              & 64.22\tiny{$\pm$6.511}\normalsize          & 79.58\tiny{$\pm$7.098}\normalsize          & 67.50\tiny{$\pm$6.778}\normalsize             & 94.31\tiny{$\pm$3.240}\normalsize          & 89.69\tiny{$\pm$3.656}\normalsize              & 50.90\tiny{$\pm$0.368}\normalsize              & 79.83\tiny{$\pm$6.519}\normalsize                  & 46.81\tiny{$\pm$2.272}\normalsize              & 81.86\tiny{$\pm$5.421}\normalsize                  & 87.35\tiny{$\pm$3.301}\normalsize          \\
GraphSAGE                        & \underline{96.18\tiny{$\pm$1.170}}\normalsize          & \underline{96.53\tiny{$\pm$0.876}}\normalsize        & \underline{96.34\tiny{$\pm$0.547}}\normalsize              & \underline{99.66\tiny{$\pm$0.052}}\normalsize       & 99.77\tiny{$\pm$0.111}\normalsize              & 53.66\tiny{$\pm$4.420}\normalsize              & 76.22\tiny{$\pm$13.96}\normalsize                  & 52.24\tiny{$\pm$3.378}\normalsize              & 92.30\tiny{$\pm$5.397}\normalsize                  & 90.47\tiny{$\pm$4.891}\normalsize          \\
\hline
GCN-ts                           & 57.54\tiny{$\pm$2.501}\normalsize          & 70.64\tiny{$\pm$5.120}\normalsize          & 60.17\tiny{$\pm$3.502}\normalsize          & 92.81\tiny{$\pm$1.751}\normalsize          & 87.94\tiny{$\pm$2.951}\normalsize          & 77.36\tiny{$\pm$2.866}\normalsize          & 88.91\tiny{$\pm$2.005}\normalsize          & 81.86\tiny{$\pm$1.906}\normalsize          & 97.94\tiny{$\pm$0.435}\normalsize          & 97.08\tiny{$\pm$0.195}\normalsize           \\
GAT-ts                           & 87.34\tiny{$\pm$16.42}\normalsize          & 92.09\tiny{$\pm$11.27}\normalsize          & 86.15\tiny{$\pm$21.60}\normalsize          & 92.66\tiny{$\pm$19.50}\normalsize          & 97.83\tiny{$\pm$2.878}\normalsize          & 75.90\tiny{$\pm$8.216}\normalsize          & 83.30\tiny{$\pm$5.774}\normalsize          & 77.98\tiny{$\pm$5.691}\normalsize          & 97.41\tiny{$\pm$1.258}\normalsize          & 96.09\tiny{$\pm$1.467}\normalsize           \\
GraphSAGE-ts                     & 95.26\tiny{$\pm$1.434}\normalsize          & 96.51\tiny{$\pm$1.347}\normalsize          & 95.84\tiny{$\pm$0.660}\normalsize          & 99.61\tiny{$\pm$0.063}\normalsize          & \underline{99.83\tiny{$\pm$0.090}}\normalsize           & \underline{95.69\tiny{$\pm$1.674}}\normalsize          & \underline{99.04\tiny{$\pm$0.951}}\normalsize          & \underline{97.27\tiny{$\pm$0.716}}\normalsize          & \underline{99.74\tiny{$\pm$0.075}}\normalsize          & \underline{99.94\tiny{$\pm$0.012}}\normalsize           \\
\hline
LMTracker                        & 65.88\tiny{$\pm$1.829}\normalsize          & 93.88\tiny{$\pm$0.955}\normalsize          & 72.08\tiny{$\pm$2.373}\normalsize          & 93.51\tiny{$\pm$1.085}\normalsize          & 95.67\tiny{$\pm$0.639}\normalsize          & 63.25\tiny{$\pm$1.999}\normalsize          & 80.88\tiny{$\pm$2.980}\normalsize          & 63.06\tiny{$\pm$3.893}\normalsize          & 74.35\tiny{$\pm$4.595}\normalsize          & 79.82\tiny{$\pm$3.856}\normalsize           \\
Euler GCN-GRU                    & 50.26\tiny{$\pm$0.013}\normalsize          & 94.82\tiny{$\pm$0.520}\normalsize          & 50.37\tiny{$\pm$0.033}\normalsize          & 99.39\tiny{$\pm$0.035}\normalsize          & 99.04\tiny{$\pm$0.187}\normalsize          & -                        & -                        & -                        & -                        & -                         \\
Euler GCN-LSTM                   & 50.28\tiny{$\pm$0.012}\normalsize          & 94.33\tiny{$\pm$0.278}\normalsize          & 50.41\tiny{$\pm$0.027}\normalsize          & 99.42\tiny{$\pm$0.023}\normalsize          & 98.60\tiny{$\pm$0.263}\normalsize          & -                        & -                        & -                        & -                        & -                         \\
\hline
\textbf{LMDetect(ours)}          & \textbf{98.20\tiny{$\pm$0.819}\normalsize} & \textbf{99.89\tiny{$\pm$0.194}\normalsize} & \textbf{99.03\tiny{$\pm$0.467}\normalsize} & \textbf{99.91\tiny{$\pm$0.044}\normalsize} & \textbf{99.99\tiny{$\pm$0.004}\normalsize} & \textbf{97.46\tiny{$\pm$0.506}\normalsize} & \textbf{99.46\tiny{$\pm$0.468}\normalsize} & \textbf{98.43\tiny{$\pm$0.324}\normalsize} & \textbf{99.87\tiny{$\pm$0.026}\normalsize} & \textbf{99.99\tiny{$\pm$0.004}\normalsize}  \\
\hline\hline
\end{tabular}
}
\end{table*}

\subsection{Evaluation on Lateral Movement Detecttion}
% 我们在两个数据集上对不同横向移动检测方法的效果进行了评估，其结果如表所示。我们可以很明显的看到，LMDetect在两个数据集的各项检测指标上均取得了最优表现。其中，LMDetect在LANL数据集上表现尤为优异，各项指标均接近完美，精确率、召回率、F1分数、准确率和AUC均达到99%左右。在CERT数据集上，LMDetect同样表现出色，达到了97%-99%的高分。这表明，LMDetect方法具有极高的鲁棒性和泛化能力，能够在不同数据集上实现稳定的性能优势。
We evaluate the effectiveness of different lateral movement detection methods across two datasets, with the results presented in Table~\ref{tab:baseline-results}. It is evident that \model outperforms other methods across all detection metrics on both datasets. Notably, \model performes exceptionally well on LANL dataset, achieving near-perfect recall, F1 score, accuracy, and AUC at approximately 99\%. On CERT dataset as well, \model also demonstrates outstanding performance with high scores. These results indicate that \model possesses strong robustness and generalization, consistently achieving superior performance across various datasets.

% 其次，GNN-based方法（GCN, GAT, GraphSAGE）的整体表现不如LMDetect。在LANL数据集中，虽然GraphSAGE相对表现较好，但仍未达到LMDetect的水平。而GAT和GCN则进一步落后，尤其是在precision和recall指标上表现不佳,说明存在较大的误报率和错报率。在CERT数据集中，这几种方法的表现普遍下降，表明其在该数据集上的适用性受到限制。究其原因，GCN和GAT在刻画实体特征时会利用周围所有信息，因此无法有效避免日志中噪声数据的影响，而GraphSAGE会随机采样周围的信息进行利用，一定程度上缓解噪声的干扰，防止模型的过拟合。
Moreover, the GNN-based methods (GCN, GAT, GraphSAGE) underperform compared to \model. While GraphSAGE performes relatively better on LANL dataset, it still does not reach the level of \model. GAT and GCN lag even further behind, particularly struggling in precision and recall, indicating high false positive and false negative rates. On CERT dataset, the performance of these methods generally declines, highlighting their limited applicability on this dataset. The underlying reason is that GCN and GAT rely on all surrounding information when capturing entity features, making them less effective at alleviating the impact of noisy data in logs. In contrast, GraphSAGE randomly samples neighboring information, somewhat alleviating noise interference and preventing model overfitting.

% 值得注意的是，利用时间感知子图的GNN-based方法（GCN-ts, GAT-ts, GraphSAGE-ts）在大多数指标上优于使用随机采样子图的版本,尤其在CERT数据集上的表现显著提升。这表明时间感知子图比传统的随机采样子图具有更高的事件相关性，可以捕捉到与横向移动行为高度关联的上下文信息，进一步增强检测精度。
Notably, the GNN-based methods that use time-aware subgraphs (GCN-ts, GAT-ts, GraphSAGE-ts) outperform their random-sampling counterparts across most metrics, with particularly significant improvement on CERT dataset. This suggests that time-aware subgraphs capture higher event relevance than random sampling, enabling the model to capture contextual information closely associated with lateral movement behaviors and thereby enhancing detection accuracy further.

\begin{table}[!htb] 
    \renewcommand\arraystretch{1.3} 
    \centering                 
    \caption{Ablation Experiment Results.} 
    \label{tab:Ablation}                                 
    \resizebox{\linewidth}{!}{
\begin{tabular}{c|ccccc} 
\hline\hline
\multirow{2}{*}{\textbf{Method}}  & \multicolumn{5}{c}{\textbf{LANL}}                                                                                                                                             \\ 
\cline{2-6}
                                  & \textbf{\textbf{\textbf{\textbf{Precision}}}} & \textbf{\textbf{\textbf{\textbf{Recall}}}} & \textbf{\textbf{\textbf{\textbf{F1}}}} & \textbf{\textbf{\textbf{\textbf{Accuracy}}}} & \textbf{\textbf{\textbf{\textbf{AUC}}}}  \\ 
\hline
\textbf{\textbf{LMDetect w/o tg}} & \textbf{98.66\tiny{$\pm$0.504}\normalsize}     & 98.59\tiny{$\pm$0.365}\normalsize                     & 98.62\tiny{$\pm$0.140}\normalsize                            & 99.87\tiny{$\pm$0.013}\normalsize                                  & 99.77\tiny{$\pm$0.111}\normalsize                              \\
\textbf{\textbf{LMDetect w/o le}} & 98.24\tiny{$\pm$0.898}\normalsize              & 99.33\tiny{$\pm$0.675}\normalsize                     & 98.77\tiny{$\pm$0.473}\normalsize                            & 99.89\tiny{$\pm$0.044}\normalsize                                  & 99.99\tiny{$\pm$0.007}\normalsize                              \\
\textbf{\textbf{LMDetect w/o ge}} & 85.74\tiny{$\pm$14.39}\normalsize              & 93.37\tiny{$\pm$6.924}\normalsize                     & 87.93\tiny{$\pm$12.19}\normalsize                            & 98.25\tiny{$\pm$2.099}\normalsize                                  & 98.13\tiny{$\pm$3.170}\normalsize                              \\
\textbf{\textbf{LMDetect w/o pe}} & 98.09\tiny{$\pm$0.938}\normalsize              & 99.83\tiny{$\pm$0.261}\normalsize                     & 98.94\tiny{$\pm$0.558}\normalsize                            & 99.90\tiny{$\pm$0.053}\normalsize                                  & 99.99\tiny{$\pm$0.008}\normalsize                              \\
\textbf{\textbf{LMDetect}}        & 98.20\tiny{$\pm$0.819}\normalsize              & \textbf{\textbf{99.89\tiny{$\pm$0.194}\normalsize}}   & \textbf{\textbf{99.03\tiny{$\pm$0.467}\normalsize}}          & \textbf{\textbf{99.91\tiny{$\pm$0.044}\normalsize}}               & \textbf{\textbf{99.99\tiny{$\pm$0.004}\normalsize}}           \\
\hline\hline
\multirow{2}{*}{\textbf{Method}}  & \multicolumn{5}{c}{\textbf{CERT}}                                                                                                                                             \\ 
\cline{2-6}
                                  & \textbf{\textbf{\textbf{\textbf{Precision}}}} & \textbf{\textbf{\textbf{\textbf{Recall}}}} & \textbf{\textbf{\textbf{\textbf{F1}}}} & \textbf{\textbf{\textbf{\textbf{Accuracy}}}} & \textbf{\textbf{\textbf{\textbf{AUC}}}}  \\ 
\hline
\textbf{\textbf{LMDetect w/o tg}} & 67.37\tiny{$\pm$1.823}\normalsize          & 96.45\tiny{$\pm$4.801}\normalsize                     & 79.33\tiny{$\pm$2.450}                     & 99.16\tiny{$\pm$0.117}          & 99.65\tiny{$\pm$0.232}                      \\
\textbf{\textbf{LMDetect w/o le}} & 97.15\tiny{$\pm$0.028}\normalsize          & \multicolumn{1}{l}{99.28\tiny{$\pm$0.472}} & \multicolumn{1}{l}{97.66\tiny{$\pm$0.243}} & \multicolumn{1}{l}{99.68\tiny{$\pm$0.022}} & 99.86\tiny{$\pm$0.002}                      \\
\textbf{\textbf{LMDetect w/o ge}} & 75.90\tiny{$\pm$8.216}\normalsize          & 83.30\tiny{$\pm$5.774}                     & 79.87\tiny{$\pm$5.691}                     & 97.41\tiny{$\pm$1.258}                     & 96.09\tiny{$\pm$1.467}                      \\
\textbf{\textbf{LMDetect w/o pe}} & 97.18\tiny{$\pm$0.236}\normalsize          & 99.25\tiny{$\pm$0.452}                     & \multicolumn{1}{l}{97.66\tiny{$\pm$0.151}} & \multicolumn{1}{l}{99.81\tiny{$\pm$0.017}} & \multicolumn{1}{l}{99.85\tiny{$\pm$0.016}}  \\
\textbf{\textbf{LMDetect}}        & \textbf{97.46\tiny{$\pm$0.503}\normalsize} & \textbf{99.46\tiny{$\pm$0.467}}            & \textbf{98.44\tiny{$\pm$0.300}}            & \textbf{99.85\tiny{$\pm$0.028}}            & \textbf{99.95\tiny{$\pm$0.012}}            \\
\hline\hline                         
\end{tabular}}
\end{table}

% 此外LMTracker和Euler在召回率方面的表现相对较好，但精确率较低，也表现出较大的误报率。究其原因，这两种方法通过链路预测的方式来进行横向移动检测，孤立地看待网络行为，忽略了横向移动通常涉及一系列连续行为的事实，因此检测性能存在局限。
Additionally, LMTracker and Euler perform relatively well in recall but demonstrate lower precision, along with a higher false positive rate. This limitation arises because they rely on link prediction for lateral movement detection, viewing network behavior in isolation and overlooking the fact that lateral movement typically involves a series of continuous actions, which restricts detection performance.

% 相比之下，我们的LMDetect利用时间感知子图生成模块来抽取验证事件周围的高相关性信息并减少噪声，同时利用多尺度注意力编码模块全面表征横向移动攻击者的行为模式，在横向移动检测中展现出极高的检出率和极低的误报率、错报率。
In contrast, our \model employs a time-aware subgraph generation module to extract highly relevant contextual information around authentication events and reduce noise. Furthermore, it uses a multi-scale attention encoding module to comprehensively characterize the behavior patterns of lateral movement attackers, achieving an exceptionally high detection rate with minimal false positive and false negative rates in lateral movement detection.

\subsection{Ablation Analysis}
% 为了评估 LMDetect 中各模块的贡献，我们设计了消融实验，其结果如表所示。具体而言，‘LMDetect w/o tg’ 表示在去除时间感知子图生成模块后，利用随机采样的子图进行横向移动检测。‘LMDetect w/o le’、‘LMDetect w/o pe ’和 ‘LMDetect w/o ge’ 则分别表示去除局部注意力编码模块、相对位置编码模块和全局注意力编码模块后的横向移动检测。
% 我们可以得出以下结论：1）时间感知子图模块可以更好地捕捉和目标认证事件高度相关的上下文信息来辅助横向移动检测，而随机采样在一定程度上会引入噪声。该模块的优势也能从表5中GNN和GNN-ts效果的对比中体现；2）去掉局部注意力模块后，模型的错报率有所提升，整体性能略微下降，表明该模块可以略微提升模型对实体的表征能力；3）去掉全局注意力模块后，模型的性能明显降低，表明该模块可以有效地捕捉认证事件的高阶交互特征和长距离依赖，显著增强横向移动检测；4）去掉相对位置编码模块后，模型的性能略微下降，表明该模块可以在一定程度上模拟认证事件的行为演化，捕捉长距离依赖。
% 整体而言，LMDetect的各个模块设计从不同角度出发提供不同程度的贡献，全面提升横向移动检测性能。
To evaluate the contribution of each module in \model, we conduct ablation experiments, with results shown in Table~\ref{tab:Ablation}. Specifically, ``\model w/o tg'' represents lateral movement detection using randomly sampled subgraphs after removing the time-aware subgraph generation module. ``\model w/o le'', ``\model w/o pe'' and ``\model w/o ge'' represent the removal of the local attention encoding module, relative position encoding module, and global attention encoding module, respectively.

From the results, we can draw the following conclusions: 
1) The time-aware subgraph generation module can effectively capture contextual information highly relevant to the target authentication event, aiding in lateral movement detection, whereas random sampling tends to introduce noise. This module's advantage is also evident in the performance comparison between GNN and GNN-ts in Table~\ref{tab:baseline-results}; 
2) Removing the local attention encoding module slightly increases the false positive rate, with a minor drop in overall performance, suggesting that this module marginally enhances the model's capability to characterize entity features; 
3) Performance significantly declines when the global attention encoding module is removed, indicating that this module effectively captures high-order interaction features and long-distance dependencies in authentication events, substantially enhancing lateral movement detection; 
\begin{figure}[!htb]
    \centering
    \includegraphics[width=0.9\linewidth]{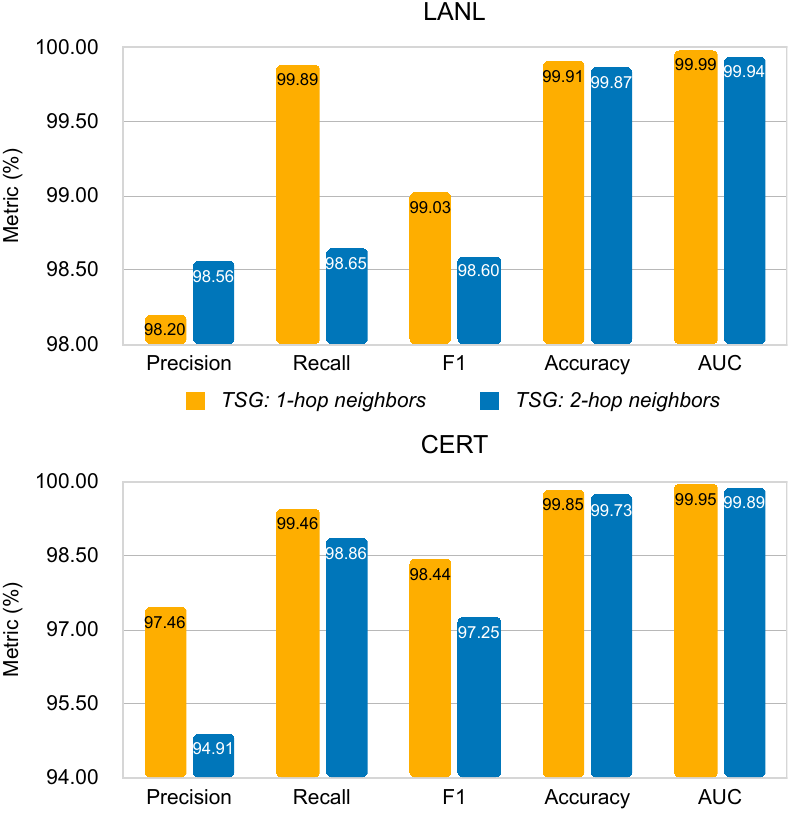}
    \caption{Impact of time-aware subgraph on detection performance.}
    \label{fig:TSG_k}
\end{figure}
4) Removing the relative position encoding module leads to a slight performance decrease, suggesting that this module simulates the behavioral evolution of authentication events to some extent and captures long-range dependencies.

Overall, the design of each module in \model contributes from different perspectives, collectively enhancing the performance of lateral movement detection.

\subsection{Analysis of Time-aware Subgraph}
% 我们进一步分析了时间感知子图规模对横向移动检测的影响。具体而言，我们在本文中仅对目标事件中所含实体的一跳邻居进行过滤来得到时间感知子图，表示为TSG: 1-hop neighbors。我们进一步对1跳和2跳邻居进行过滤来构建相似规模的时间感知子图，表示为TSG: 2-hop neighbors。我们分别利用这两种时间感知子图来进行横向移动检测，其结果如图所示，从中我们可以明显地看到，利用仅包含一跳信息的时间感知子图来进行横向移动检测可以取得更好的效果。这个现象表明，利用直接交互的1跳信息来辅助横向移动检测通常更有效，而引入更多间接交互的2跳信息的同时也往往会引入更多的噪声，使得模型对横向移动行为的检测更加保守，提高了漏报率，削弱了整体性能。
We further analyze the impact of time-aware subgraph on lateral movement detection. Specifically, in this work, we filter only the 1-hop neighbors of entities within the target event to obtain the time-aware subgraph, denoted as ``TSG: 1-hop neighbors''. Additionally, we construct similarly scaled time-aware subgraphs by filtering both 1-hop and 2-hop neighbors, denoted as ``TSG: 2-hop neighbors''. 
We use these two types of time-aware subgraphs for lateral movement detection, with the results shown in Figure~\ref{fig:TSG_k}. It is clear that using the time-aware subgraph containing only 1-hop information yields better results for lateral movement detection. This finding suggests that leveraging 1-hop information from direct interactions is generally more effective for aiding lateral movement detection, while incorporating 2-hop information from indirect interactions often introduces more noise. This added noise makes the model more conservative in detecting lateral movement, increasing the false negative rate and diminishing overall performance.

% We further analyze the effect of the sampling order $k$ in TSG on the performance of \textit{LMDetect} detection. In the experiments, $k$ is set to 1, 2 respectively. we ensure that the two sets of experiments can generate similar subgraph sizes. Fig~\ref{fig:TSG_k} demonstrates the comparison of the performance results of the two sets of experiments. 

% When \textit{LMDetect} samples first-order neighbors for target authentication events using TSG, the detection performance outperforms the results of sampling second-order neighbors in most metrics. We argue that sampling second-order neighbors introduces more node information into the model, which, while capturing a wider range of graph structures, also increases the complexity of the subgraph, making it more difficult for the model to process and learn. In lateral movement detection, direct neighbors may more directly reflect patterns of malicious behavior. Thus, sampling first-order neighbors allows the model to focus more on these key features, resulting in better performance on metrics such as Recall, F1-score, accuracy, and AUC. While sampling second-order neighbors can slightly improve Precision, this improvement may come at the cost of other metrics, as the model may miss some malicious behaviors when it is more careful in determining events as benign, leading to a decrease in other metrics. In future work, we will further explore the optimization of sampling higher-order neighbors.

\begin{figure}[!htb]
    \centering
    \includegraphics[width=\linewidth]{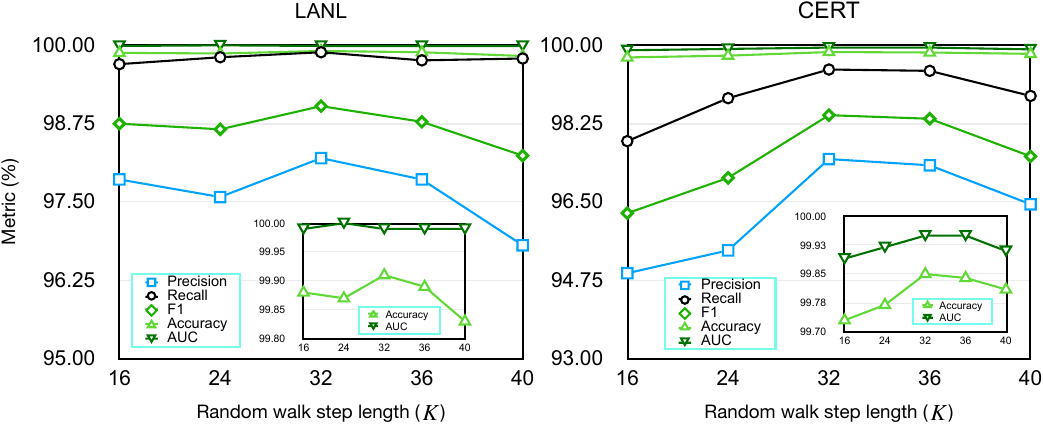}
    \caption{Impact of random walk step length in relative position encoding.}
    \label{fig:k-steps}
\end{figure}
\subsection{Analysis of Relative Position Encoding}
% 我们进一步分析相对位置编码中的关键参数设置（即随机游走步长$K$）对model检测性能的影响。具体而言，我们变化K在{16，24，32，36，40}中，并分析模型的性能波动，如图所示，从中我们可以看到，随着K的增大，模型性能在整体上呈现先上升后下降的趋势，其中当K=32时，模型达到了最优的检测性能。究其原因，相对位置编码模块通过模拟随机游走来捕捉认证子图中各实体之间的相对依赖关系，以此来强化模型对认证子图结构的理解。当步长较小时，模型仅能捕捉到实体间的局部结构，限制了对长距离依赖的刻画；当步长较大时，相对位置编码包含了过多的噪声和冗余信息，在一定程度上干扰模型的决策。因此，合适的步长选择与时间感知子图的规模密切相关，反映了模型在捕捉局部和全局依赖上的权衡。
We further analyze the influence of key parameter in relative position encoding (i.e., random walk step length $K$) on model's detection performance. Specifically, we varies $K$ within the set \{16, 24, 32, 36, 40\} and observe the performance fluctuations, as shown in Figure~\ref{fig:k-steps}. We observe that as $K$ increases, the model's performance initially improves but then declines, reaching optimal detection performance when $K = 32$.
The underlying reason is that the relative position encoding module captures the relative dependencies between entities in the authentication subgraph by simulating random walks, thereby enhancing the model's understanding of the subgraph structure. When the step length is small, \model only captures the local structure between entities, limiting its ability to represent long-range dependencies. Conversely, when the step length is too large, the relative position encoding introduces excessive noise and redundant information, which interferes with the model's reasoning to some extent. Therefore, an appropriate choice of step length is closely related to the scale of the subgraph, reflecting a balance in the model's ability to capture both local and global dependencies.

\section{Conclusion}
% Nowadays, lateral movement is increasingly becoming a focus of attention in the field of cybersecurity. In this paper, we provide a new perspective for lateral movement detection and propose an end-to-end multi-scale lateral movement detection framework called \textit{LMDetect}, which focuses on graph depth analysis of host authentication log data and integrates a time-aware subgraph generation module and a multi-scale attention mechanism to efficiently detect the lateral movement behavior of attackers. Our extensive experiments on multiple log datasets demonstrate the state-of-the-art detection performance and excellent generalization capability of our framework. Moreover, our framework has good portability, e.g., for other APT attack technique detection platforms, which will be discussed in future work.

In this work, our proposed \model framework effectively addresses the challenges of lateral movement detection in network authentication logs. By modeling authentication logs as a heterogeneous authentication multigraph and employing a time-aware subgraph classification approach, \model captures both local and global dependencies, thus enhancing the detection performance for lateral movements. The use of a time-aware subgraph generator ensures that the model can efficiently extract relevant contextual information, while the multi-scale attention encoder comprehensively characterizes the behavior patterns of lateral movement. Experimental results on real-world datasets demonstrate that \model achieves state-of-the-art performance across multiple evaluation metrics, consistently outperforming existing detection approaches. These results confirm \model's robustness and scalability in detecting complex attack strategies, thereby contributing significantly to the field of network security. Future work may involve adapting the model to other types of cyber attacks, as well as exploring the applicability of time-aware subgraph classification in different security contexts.


% % use section* for acknowledgment
% \ifCLASSOPTIONcompsoc
%   % The Computer Society usually uses the plural form
%   \section*{Acknowledgments}
% \else
%   % regular IEEE prefers the singular form
% \fi




%\bibliographystyle{IEEEtran}

\bibliographystyle{IEEEtran}
\bibliography{refs,IEEEabrv} 



% \end{thebibliography}

\end{document}